\documentclass[final]{siamltex}

\usepackage[dvips]{graphicx}
\usepackage{amsmath}
\usepackage{amssymb}
\usepackage{psfrag}
\usepackage{color}
\usepackage{subfigure}
\newcommand{\bs}{\boldsymbol}
\newcommand{\bsf}{\mathbf}
\def\bxi{\bs{\xi}}
\def\bt{\bs{\theta}}
\def\RR{ \mathbb R}

\newcommand{\refeq}[1]{Equation (\ref{#1})}
\newcommand{\ee}{\end{equation}}
\newcommand{\be}{\begin{equation}}
\newcommand{\ec}{\end{center}}
\newcommand{\bc}{\begin{center}}
\newcommand{\eea}{\end{eqnarray}}
\newcommand{\bea}{\begin{eqnarray}}
\newcommand{\bd}{\begin{description}}
\newcommand{\ed}{\end{description}}
\newcommand{\bi}{\begin{itemize}}
\newcommand{\ei}{\end{itemize}}

\newcommand{\pe}{\psi}
 
\def\ds{\displaystyle} 
\def\e{{\epsilon}}

\def\fishpack{{FISHPACK}} 
 
\def\gmres{{GMRES}} 
\def\gmresm{{\rm GMRES($m$)}}

\def\bfE{\mbox{\boldmath$E$}}
\def\bfG{\mbox{\boldmath$G$}}

\title{Uncertainty quantification in complex systems using approximate solvers}

\author{Phaedon-Stelios Koutsourelakis \thanks{School of Civil and environmental Enginerring \& Center for Applied Mathematics, 369 Hollister Hall,  Cornell University, Ithaca, NY 14853, ({\tt pk285@cornell.edu}).} }

\begin{document}

\maketitle

\begin{abstract}
This paper proposes a novel uncertainty quantification framework for computationally demanding systems characterized by a large vector of non-Gaussian uncertainties. It combines state-of-the-art techniques in advanced Monte Carlo sampling with Bayesian formulations. The key departure from existing works is the use of inexpensive, approximate computational models in a rigorous manner. Such models can readily be derived by coarsening the discretization size in the solution of the governing PDEs, increasing the time step when integration of ODEs is performed, using fewer iterations if a non-linear solver is employed or making use of lower order  models. It is shown that even in cases where the inexact models provide  very poor approximations of the exact response, statistics of the latter can be quantified accurately with significant reductions in the computational effort. Multiple approximate models can be used and rigorous confidence bounds of the estimates produced are provided at all stages.

\end{abstract}

\begin{keywords} 
uncertainty quantification, Monte Carlo, Bayesian, nonparametric regression
\end{keywords}

\begin{AMS}
65C05,62F15,62G08
\end{AMS}

\pagestyle{myheadings}
\thispagestyle{plain}
\markboth{P.S. KOUTSOURELAKIS}{UNCERTAINTY QUANTIFICATION WITH APPROXIMATE SOLVERS}

\section{Introduction and examples}

Scientists have come to  recognize the stochastic aspects inherent in several physical systems and processes and seek ways to quantify the probabilistic characteristics of their behavior.
Their analysis tools are usually restricted to elaborate legacy codes which have been developed over a long period of time and are generally well-tested. They do not however  include any stochastic components and their  alteration is commonly impossible or ill-advised. In many problems of engineering or physical  interest the  only feasible solution for uncertainty quantification is provided by non-intrusive methodologies.

Traditionally, two approaches have have attracted most attention. On one hand methods based on polynomial chaos expansions (PC, \cite{wie38hom})) and on the other techniques anchored around Monte Carlo simulations. PC models, although originally developed as intrusive techniques (\cite{gha91sto}), have grown into prominence in recent years with the development of non-intrusive, stochastic collocation approaches (\cite{xiu05hig,gan07spa}). They are based on a representation of the random input by a finite number of uncorrelated random variables (usually normally distributed) and orthogonal polynomials (usually Hermite). The solution or output process is expressed with respect to the same basis  and  the coefficients of the expansion are determined by calculating  weighted residuals or using a collocation approach. Although mathematically elegant, PC-based approaches utilize a second-order matching (up to the autocovariance function) of the input processes which does not account for important higher order statistics that might affect the system's response. The computational effort grows with the number of random variables used to approximate the input which also adversely affects the accuracy, particular in the stochastic collocation version, as an interpolation in a very high dimensional space is required.

Standard Monte Carlo simulations require a minimal implementation overhead as the coupling with existing deterministic solvers is trivial. Most often than not however, in systems of physical interest, each of the runs of the forward solver  requires several  CPU-hours and multiple processors. Even though the convergence rate $O(\sqrt{N})$ (where $N$ is the number of independent samples) is independent of the dimensionality of the random input, it is sufficiently slow to constitute the method impractical or infeasible as several calls to forward solver have to be made to achieve good accuracy. Another disadvantage of classical Monte Carlo is that it does not directly provide  confidence intervals for the estimates produced as those are usually based on asymptotic results (i.e. when $N \to \infty$) and the central limit theorem (\cite{ras03bay}).
  Recent years have seen significant progress in the development of advanced Monte Carlo techniques which employ evolutionary strategies in combination with Markov Chain Monte Carlo (MCMC) and Importance Sampling (\cite{gel98sim,nea01ann,kou08des}). In many cases this has led to algorithms which require $10$ or $100$ times less samples in order to produce estimates of the same accuracy (\cite{au01est}).  

Despite this dramatic improvements, the associated computational effort can still be tremendous for systems of practical interest, where even a modest number of $100$ or $1000$ runs can be infeasible.
It is obvious that a new perspective is needed. In the author's opinion this can be achieved if analysis goes beyond the black-box solver as the only means of probing the problem of interest. Indeed in many situations, several other pieces of knowledge and structural elements of the problem at hand, are readily available  but left unexploited. For example, quite frequently the computational models of interest involve the solution of  a system of PDEs using  Finite Elements (FE) or Finite Differences (FD). These imply the spatio-temporal discretization of the governing differential equations  and quite often the mesh sizes or time steps have to be  particularly small in order to capture the salient features of the solution. Without recourse to rigorous mathematical proofs, it is well-known that an FE solver that operates on a coarser spatio-temporal grid can give an {\em approximate} solution at a lower computational cost as the system of equations that need to be solved are smaller. The deviation from  the "exact" (or reference)  solution can be significant but in principle this approximate solver can  be used to obtain some, inaccurate of course, information about our exact model. As it will be demonstrated in the sections to come, it is not important if the solutions of the approximate solver deviate significantly from the exact, but it suffices that they exhibit some sort of dependence. It is this dependence that we will exploit in a general and rigorous computational framework in conjuction with a few, carefully selected runs of the full, exact model. 
We will make no claims about the optimality of the approximate solvers selected. In fact as it will be demonstrated in the examples even crude approximations can yield impressive increases in computational efficiency. Furthermore, the framework proposed allows for the introduction of several such approximate models similar to the way one would elicit opinions from multiple experts before making a decision. In that respect even low-order, fast PC models can be utilized.
Such an approach can be employed even in  cases where  no accurate computational model exists but rather we have to rely on experiments in order to collect the necessary information about the system. Since conducting experiments can be costly and time consuming it is desirable to minimize them by making use of approximate computational models that might be available.

To that end  we investigate Bayesian alternatives to classical uncertainty quantification techniques 
In particular we formulate  a regression problem that establishes the connection between the response values from the approximate and exact solver. This is achieved using a flexible, non-parametric Bayesian model that employs   a very efficient Sequential Monte Carlo inference algorithm. An added advantage of this approach is that prior knowledge or expertise of the analyst regarding the relationship between approximate and exact solvers can be readily incorporated in the prior distributions. Once this relation is established, the posterior distribution can be readily used to obtain estimates and confidence intervals on the output statistics  of interest. These can in turn be used to actively and adaptively improve the accuracy of the regression by performing runs of the expensive solver in regions that contribute most significantly to these uncertainties. An interesting extension involves using more than one  approximate solvers  simultaneously in order to improve the accuracy and computational efficiency of the model. This resembles mixtures of experts models that are commonly used in various statistical applications, as each approximate solver provides some, generally incomplete, information about the exact model which is then aggregated in order to obtain the best possible estimate. 

\section{Proposed Approach}
Let $(\Omega, \mathcal{F},\mathcal{P})$  be a complete probability
space, where $\Omega$ is the event space, $\mathcal{F}$ the $\sigma$-algebra, and $\mathcal{P}$  the probability  measure. Consider the following stochastic differential equation:
\be
\label{eq:sde1}
\mathcal{L}(u(\bs{z}, t) ; \bxi(\omega)) = f(\bs{z},t; \bxi(\omega)), \quad \bs{z} \in \mathcal{D}, ~t>0
\ee
defined on the domain $\mathcal{D} \subset \RR^q$  ($q = 1, 2, 3$) with appropriate initial/boundary conditions which might also depend on the vector of uncertainties represented by $\bxi(\omega):\Omega \to \RR^d$. We are particularly interested in the most general and difficult case where $\bs{\xi}$ is of very large dimension (i.e. $d >> 1$ ) and it should not or  cannot be condensed using any of the standard dimension reduction techniques (e.g. PCA). Let 
 $u(\bs{z},t; \bxi( \omega))$ denote  the solution process  which satisfies  \refeq{eq:sde1} for $\mathcal{P}$-almost everywhere. We are interested in the statistics of the output itself or of a function thereof which we denote by $y(\bxi) : \RR^d \to \RR^r$ emphasizing the dependence on the vector of input uncertainties $\bxi$.  We further postulate the existence of a  forward solver of the linear/nonlinear equation in \refeq{eq:sde1} that corresponds to a deterministic version of the differential operator $\mathcal{L}$ (for fixed $\bxi$). 
In general however one might consider reduced versions of the above problem (i.e. dependence only on time or space) or even systems (and associated computational models)  which are not governed by SPDEs but nevertheless characterized by a high-dimensional vector of input uncertainties $\bxi$. For illustration  purposes we will restrict the presentation to the case that $y$ is scalar  (i.e. $r=1$). Naturally $y$ may  depend on other deterministic parameters which are omitted for economy of notation.

The input uncertainties $\bs{\xi}$ are characterized by a probability density $\pi_{\xi}$. In order for the problem to be well-posed, $\pi_{\xi}$ need not be known analytically but it suffices to be able to draw samples from $\pi_{\xi}$.  Our goal is to calculate statistics of the response, e.g.:
\be
Pr [y \in \mathcal{A}] = \int \bs{1}_{ \mathcal{A}}(y(\bs{\xi}))~ \pi_{\xi} (\bs{\xi})~d\bs{\xi}
\label{eq:def1}
\ee
where $\bs{1}_{ \mathcal{A}}$ is the indicator function of a $\pi_{\xi}$-measurable subset $\mathcal{A}$, or:
\be
E[ h(y)] =\int h(y(\bs{\xi}))~ \pi_{\xi} (\bs{\xi})~d\bs{\xi}
\label{eq:def2}
\ee
where $h$ is any $\pi_{\xi}$-integrable function.

Quite often the statistics of interest involve very rare events (i.e. $Pr [y \in  \mathcal{A}] << 1$], as is the case for example in molecular dynamics simulations where transition to a local minimum of the free-energy landscape happens infrequently or in estimating reliability of mechanical components. In other cases we are interested in expectations of multimodal functions $h$ as is the case in nonlinear dynamical systems where small perturbations in the input $\bs{\xi}$ can lead to significant differences in the (long-term) response.
Due to the large variance of the integrands in Equations (\ref{eq:def1}) or (\ref{eq:def2}), several calls to the forward solver have to be made (to  calculate the response $y$ for various $\bs{\xi}$'s)  which imply a significant or insurmountable computational burden, particularly in cases where each of these calls imply several CPU-hours on multiple processors.

To address these issues we postulate the existence of $M$  {\em approximate} forward solvers $x_m(\bs{\xi}): \RR^d \to \RR^r$, for $m=1,\ldots M$, as those discussed in the introduction and in the numerical examples to follow. Each of those provides {\em approximations} to the output of interest at a fraction of the computational cost. The latter requirement is the key in increasing the overall computational efficiency in the proposed framework, whereas the former condition can be interpreted very loosely. In fact it is acceptable that the $x_m$'s provide  very poor estimates  (i.e. $y(\bxi)-x_m(\bxi)$ is relatively large) as long as there is some statistical dependence between them. In that sense $x_m$ might not even correspond to the same output quantities, although in such cases the selection of reasonable   approximate solvers can be less straightforward. For the purposes of this work, $x_m$'s are viewed as (potentially) {\em biased} and {\em partial } predictors of the output $y$. Our goal is to quantify the information these predictors provide for the purposes  of estimating statistics of $y$ at a fraction of the computational cost. In particular, if $\bsf{x}=(x_1,x_2, \ldots, x_m)$,  \refeq{eq:def1} can be rewritten as:
\begin{eqnarray}
\label{eq:def1a}
Pr [ y  \in \mathcal{A} ] & = & E_{\bsf{x}}\left[ Pr [ y \in  \mathcal{A} \mid \bsf{x} ] \right] \nonumber \\
 & =& \int Pr[y \in \mathcal{A} \mid \bsf{x}  ] \pi_x(\bsf{x}) ~d\bsf{x} \nonumber \\
 & =& \int \left( \int \bs{1}_{\mathcal{A}}(y)~ p(y \mid x) ~dy \right) ~ \pi_x(\bsf{x}) ~d\bsf{x} 
\end{eqnarray}
and  \refeq{eq:def2}:
\begin{eqnarray}
\label{eq:def2a}
E[h(y)]& = &E_{\bsf{x}}\left[ E[h(y) \mid \bsf{x} ] \right]  \nonumber \\
        & = & \int  E[h(y) \mid \bsf{x} ] \pi_x(\bsf{x}) ~d\bsf{x}  \nonumber \\
        & =& \int \left(\int h(y)~p(y\mid \bsf{x}) ~dy\right)~  \pi_x(\bsf{x}) ~d\bsf{x}
\end{eqnarray}

Hence it is apparent that estimates of  $Pr [ y \in \mathcal{A}]$ and $E[h(y)]$ can be obtained as long as the density $\pi_x(\bsf{x})=\int \delta(\bs{x}-\bs{x}(\bxi))~\pi_{\xi}(\bxi)~ d\bxi$ and conditional $p(y \mid \bsf{x})$ are known.
Given that calls to the approximate solvers are computationally inexpensive in relative terms as it will be seen in the examples of section \ref{examples}, $\pi_x(\bsf{x})$ can be readily evaluated using direct or advanced Monte Carlo techniques as those discussed previously.
The pivotal component is the conditional density $p(y \mid \bsf{x})$ which probabilistically quantifies the information that the predictors $\bsf{x}$ carry about $y$.
\begin{figure}[htp]
     \centering
     \subfigure[Extreme scenario: Independence]{
\psfrag{y}{$y$}
\psfrag{x}{$x_m$}
          \label{fig:def1a}
           \includegraphics[width=.48\textwidth,height=6.5cm]{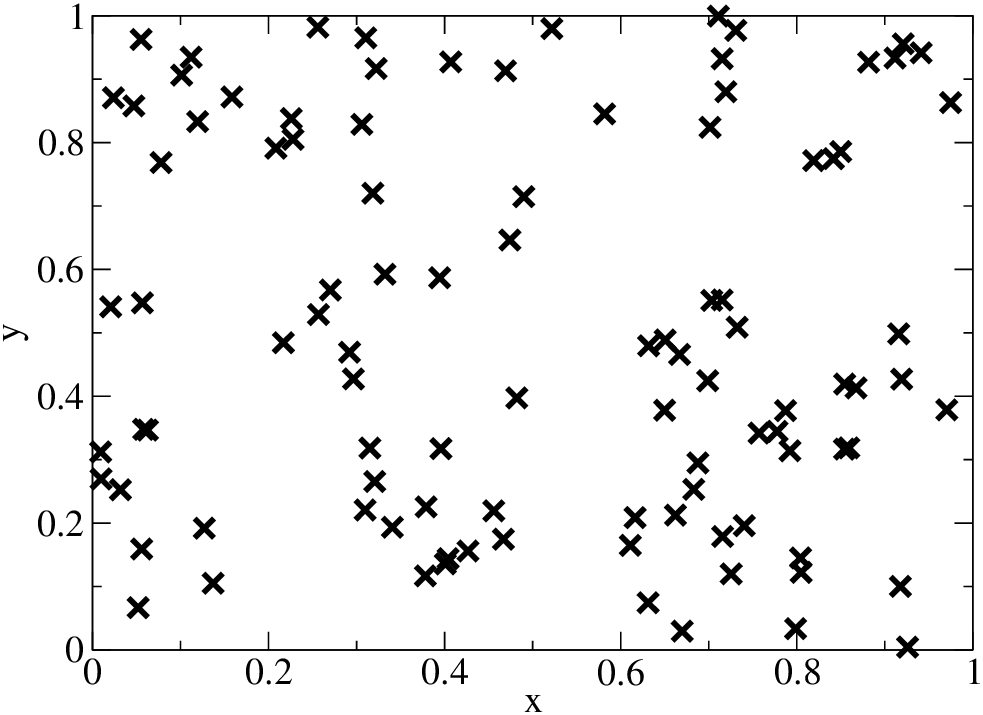}} 
\hfill
     \subfigure[Extreme scenario: One-to-one correspondence]{
\psfrag{y}{$y$}
\psfrag{x}{$x_m$}
          \label{fig:def1b}
             \includegraphics[width=.48\textwidth,height=6.5cm]{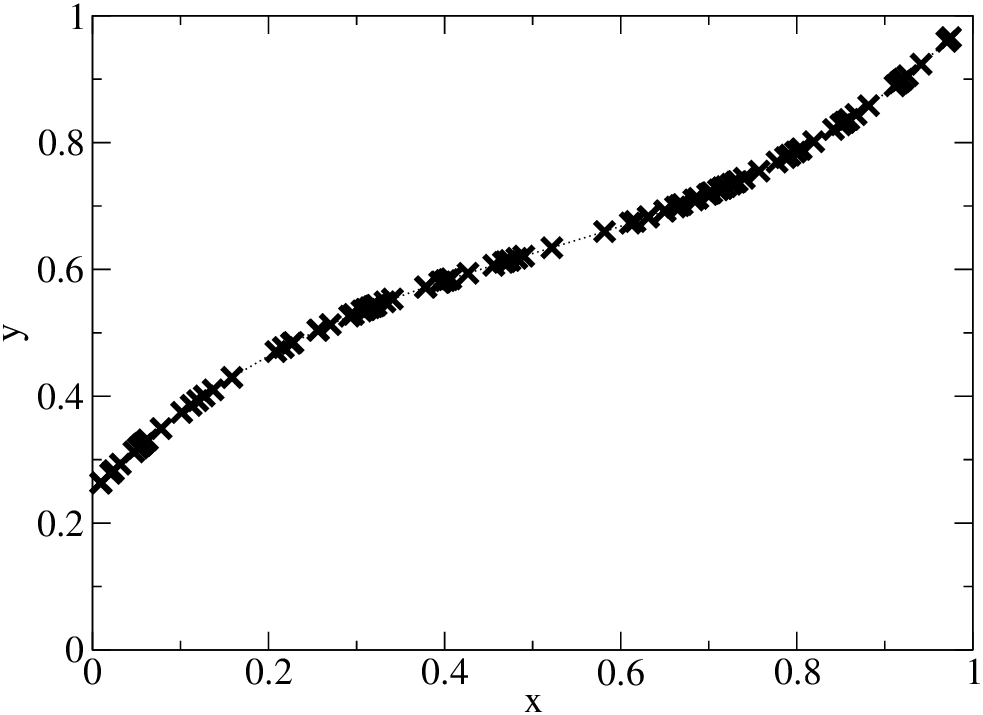}}\\
     \caption{}
     \label{fig:def1}
\end{figure}

Figure \ref{fig:def1} trivially illustrates the two extreme scenaria. On one hand $y$ and $\bsf{x}$ are statistically {\em independent}. In this case knowledge of $\bsf{x}$ is completely useless in furnishing information about $y$ and $p(y \mid \bsf{x})=p(y)$. Hence the proposed framework cannot offer any improvement. On the other extreme, there exists  an injective, deterministic  mapping  between the two quantities, i.e. $y=g(\bsf{x})$  and therefore knowing $\bsf{x}$ and its statistics translates straightforwardly to $y$ since $p(y \mid \bsf{x})=\delta (y-g(\bsf{x}))$.  The proposed methodology is applicable to all cases except the one of independence between $\bsf{x}$ and $y$.

Critical to the feasibility of proposed framework is establishing a quantitative link between the {\em exact}  $y$ and {\em approximate} $\bsf{x}$ outputs as described by $p(y \mid \bsf{x})$. This will be accomplished using computationally generated data that consist of pairs of $ \{ (\bsf{x}_i=\bsf{x}(\bxi_i),~y_i=y(\bxi_i) )\}_{i=1}^n$ obtained by running approximate and exact solvers for the same $\bxi_i$.  This is discussed in detail in the next 3 sub-sections. The task of utilizing the inferred models for the purposes of estimating \refeq{eq:def1a} or \refeq{eq:def2a} is discussed in sub-section \ref{prediction} and illustrated in the examples of section \ref{examples}.

\subsection{Hierarchical Bayesian model}
We assume that the data have been rescaled so that 
 $\bsf{x}_i \in [0,1]^M$ and consider regression models of the form:
\be
\label{eq:def3}
 y(\bxi_i)=y_i = f(\bsf{x}(\bxi_i); \bt) + \sigma~ Z_i
\ee
where $f$ is a function of the predictors $\bsf{x}=(x_1,\ldots x_M)$ and model parameters $\bt$, and $Z_i$ are i.i.d standard normal variates i.e. $Z_i \sim \mathcal{N}(0,1)$ (if $y \in \RR^r$ then $f, Z_i \in \RR^r$). \refeq{eq:def3} postulates that, given the model parameters $\bt$, for an input $\bxi_i$ for which the outputs of the approximate models are $x_m(\bxi_i)$, the target response $y(\bxi_i)$ is normally distributed with mean $f(\bsf{x}(\bxi_i); \bt)$ and standard deviation $\sigma$, i.e.:
\be
\label{eq:def4}
y_i \mid \bsf{x}(\bxi_i), \bt, \sigma \sim \mathcal{N}( f(\bsf{x}(\bxi_i); \bt), \sigma^2 \bs{I})
\ee
At first glance such a model seems highly restrictive as it is unlikely that $y$ is normally distributed (given $\bsf{x}$ and model parameters $\bt$). For that purpose we adopt a Bayesian formulation in which the model parameters $\bs{\theta}$  are assumed random and equipped with a distribution. This allows us to actually formulate a {\em family} of such models (each corresponding to a particular $\bs{\theta}$) and even though conditionally on $\bs{\theta}$, $y$ is normally distributed,  marginally (when $\bt$ are integrated out) non-Gaussian distributions can be considered. 

Bayesian formulations  differ from classical statistical approaches (frequentist) in that all unknown parameters are treated as random. Hence the results of the inference process  are not point estimates but distribution functions. 
The basic elements of Bayesian models are the {\em likelihood} function $L(\bs{\theta})=p(data \mid  \bs{\theta})$ which is a conditional probability distribution and gives a (relative) measure of the propensity of observing data  for a given model configuration specified by the parameters $\bs{\theta}$. The likelihood function is also encountered in frequentist formulations where the unknown model parameters $\bs{\theta}$ are determined by maximizing $L(\bs{\theta})$. This could be thought as the probabilistic equivalent of deterministic optimization techniques commonly used in such problems.                  The second component of Bayesian formulations  is the {\em prior} distribution $p(\bs{\theta})$ which encapsulates in a probabilistic manner any knowledge/information/insight that is available to the analyst prior to observing the data. Although the prior is a point of frequent criticism due to its
inherently subjective nature, it can prove extremely useful in the context of problems examined as it provides a mathematically consistent vehicle for  injecting the analyst's insight (whenever it is available) with regards to the relation between the exact and approximate models.                                                                                                                  The combination of {\em prior} and {\em likelihood} based on Bayes' rule yields the {\em posterior} distribution $\pi(\bs{\theta})$ which probabilistically summarizes the information extracted from the data with regards to the  unknown $\bs{\theta}$ : %
                                                                               \be                                                                                  \label{eq:m0}                                                                        \pi(\bs{\theta})=p( \bs{\theta} \mid data)=\frac{p(data \mid  \bs{\theta})~ p(\bs{\theta}) }{ p(data) } \propto p(data \mid  \bs{\theta})~ p(\bs{\theta})
\ee                                                                                                                                                             Hence Bayesian formulations allow for the possibility of multiple solutions - in fact any  $\bs{\theta}$ in the support of the likelihood and the  prior is admissible - whose  {\em relative plausibility} is quantified by the posterior. Credible or confidence intervals can be readily estimated from the posterior which quantify inferential uncertainties about the unknowns.

The crucial ingredient is of course the prior specification, not only in terms of the functional form of $p(\bt)$ but primarily in terms of the structural characteristics of the relation between $y$ and $\bsf{x}$ that is implied in \refeq{eq:def3}. It is easily understood, that any parameterization that depends on a finite number of $\bt$ will be restrictive no matter how large the family of models that it contains. Furthermore,  in order to be consistent with the {\em principle of parsimony},  prior models should make as few assumptions as possible and allow their complexity to be inferred from the data.  To satisfy the aforementioned desiderata and overcome the shortcomings of existing approaches, we propose the use of {\em nonparametric priors} (\cite{tip01spa,lia06non}). As the term can be misleading, we note that this does not imply lack of parameters but rather that the number of parameters is not {\em a priori fixed} and can change as the data dictates.
At the core of such representations, lie simple basis functions,  whose shape and location are controlled by a few parameters. The key unknown is the {\em cardinality } of the model, i.e. the number of such terms that are needed to provide a good interpretation of the data. Consider the expansion:
\vspace{-.3cm}
\be
\vspace{-.2cm}
\label{eq:m2}
{f}(\bs{x} ; \bt) = a_0 + \sum_{j=1}^k a_j K_j(\bs{x}; \bs{\phi}_j) \quad x \in D
\vspace{-.25cm}
\ee
 where $\bt=(k,\{ \bs{\phi}_j \})$, $K_j$ are {\em kernels} that serve as the basis functions of our representation and $\bs{\phi}_j$ associated parameters. Expression (\ref{eq:m2}) is motivated by the representer theorem of Kimeldorf and Wahba (\cite{kim71cor}), which states that the solution to the problem of minimizing a goodness-of-fit loss function subject to a Reproducing Kernel Hilbert Space  norm penalty lies in a subspace  represented as in \refeq{eq:m2}.
 Overcomplete representations as in \refeq{eq:m2} have
been advocated because they have greater robustness in the presence of
noise, can be sparser, and can have greater flexibility in matching structure
in the data (\cite{lew00lea,and01rob,lia06non}). One possible selection for the functional form of $K_j$, that also has an  intuitive parameterization, is isotropic, Gaussian kernels:
\be
\label{eq:m3}
K( \bs{x} ; \bs{\phi}_j)=(\bs{x}_j, \tau_j)) = \exp \{ - \tau_{j} \parallel \bs{x} -\bs{\nu}_j \parallel^2 \}
\vspace{-.25cm}
\ee
The parameters $\tau_j$  directly correspond to the scale of variability of ${f}(\bs{x})$. Large $\tau_j$'s imply narrowly concentrated  fluctuations and large values slower varying fields. The center of each kernel is specified by the location parameter $\bs{\nu}_j$. 

The parameters of the prior model adopted consist of:
\bi
\item $k$: the number of kernel functions needed,
\item $\{a_j\}_{j=0}^k$, the coefficients of the expansion  in \refeq{eq:m2}. Each of those can be a scalar or vector depending on the dimensionality of the {\em  exact} output $y$. 
\item $\{ \tau_j \}_{j=1}^k$ the precision parameters of each kernel which pertain to the scale of the unknown field(s), and
\item $\{ \bs{\nu}_j \}_{j=1}^k$ the center locations of the kernels which are points in $[0,1]^M$.
\ei

Let $\bs{\theta}_k=\{\{a_j\}_{j=0}^k,\{ \tau_j \}_{j=1}^k, \{ \bs{\nu}_j \}_{j=1}^k\}\in \bs{\Theta}_k$  denote the vector containing all the unknown  parameters and $\bt=(k,\bt_k)$. If $k$ is also assumed unknown and allowed to vary, then the dimension of $\bt_k$ is variable as well and $\bs{\Theta}_k\triangleq (\RR^{k+1})^r \times (\RR^+)^k \times ([0,1]^M)^k$.  
 For example, in the case of two approximate solvers $(x_1,x_2)$ ($M=2$)  and  a scalar $y$ ($r=1$),  $\bs{\theta}_k$ is of dimension $(k+1)+k+(2k)=1+4k$, i.e. $\bs{\Theta}_k \triangleq \RR^{1+k}\times (\RR^+)^k \times [0,1]^{2k}$. 
In accordance with the Bayesian paradigm, all unknowns are considered random and are assigned  prior distributions which quantify any information, knowledge, physical insight, mathematical constraints  that is available to the analyst before the data is processed. Naturally, if specific  information  about the relation between exact $y$  and approximate $\bs{x}$  outputs  is available it can be reflected on the prior distributions. We consider prior distributions of the following form (excluding hyperparameters):
\begin{eqnarray}
\label{eq:m10}
p(k, \{a_j\}_{j=0}^k,\{\tau_j\}_{j=1}^k, \{\bs{x}_j\}_{j=1}^k) & \propto &  p(k)   \nonumber \\ 
& \times & p( \{a_j\}_{j=0}^k \mid k )  \nonumber \\
& \times &  p(\{\tau_j\}_{j=1}^k \mid k)  \nonumber \\
& \times & p( \{\bs{x}_j\}_{j=1}^k) )    
\end{eqnarray}

 In order to increase the robustness of the model and exploit structural dependence we adopt a hierarchical prior model (\cite{gel03bay}).

\subsection{Prior Distribution}
\label{prior}
Pivotal to the robustness and expressivity  of the model  is the selection of the model size, i.e. of the number of kernel functions $k$ in \refeq{eq:m2}. This number is unknown a priori and in the absence of specific information, {\em sparse} representations should be favored.  This is not only advantageous for computational purposes, as the number of unknown parameters is proportional to $k$, but also consistent with the parsimony of explanation principle or Occam's razor (\cite{jef92ock,ras01occ,mur05not}). For that purpose, we propose  a Poisson prior for $k$:
\be
\label{eq:m5}
p(k \mid \lambda) = e^{-\lambda} \frac{\lambda^k}{k!} \qquad k=0,1,\ldots, \infty
\ee
For computational purposes, the aforementioned distribution is truncated  beyond  $k_{max}$. The latter is selected based on computational limitations and defines the support of the prior. This prior allows for representations of various cardinalities to be assessed  simultaneously with respect to the data. As a result the number of unknowns is not  fixed and the corresponding posterior has support on spaces of different dimensions as discussed in more detail in the sequence. 
 In this work, an exponential  hyper-prior is used for the hyper-parameter $\lambda$ to allow for greater flexibility and robustness i.e. $p(\lambda \mid s) =s~\exp\{-\lambda ~s \}$. After integrating out $\lambda$ we obtain:
\be
\label{eq:m5a}
p(k \mid s) \propto \frac{1}{(s+1)^{k+1}} ,  \qquad for ~ k=0,1,\ldots, k_{max}
\ee

The  parameters  $\{ \tau_j\}_{j=1}^k$  control the  scale of variability in the relation between $y$ and $\bsf{x}$. If prior information about this is available then it can be readily accounted for by appropriate prior specification. In the absence of such information however  multiple possibilities exist. 
We assumed  $\tau_j$ are a priori independent i.e. $p(\{ \tau_j\}_{j=1}^k)= \prod_{j=1}^k  p(\tau_j)$ and a  $Gamma(a_{\tau},b_{\tau})$ prior was used for each $\tau_j$:
\be
\label{eq:m9}
p(\{\tau_j\}_{j=1}^k \mid k, ~a_{\tau},b_{\tau} ) = \prod_{j=1}^k \frac{b_{\tau}^{a_{\tau}} }{\Gamma(a_{\tau}) } \tau_j^{ a_{\tau}-1 } ~\exp(-b_{\tau} \tau_j)
\ee
This has a mean $a_{\tau} / b_{\tau}$ and coefficient of variation $1/\sqrt{a_{\tau}}$. Diffuse versions  can be adopted  by selecting small $a_{\tau}$.
 A non-informative prior $p(\tau_j)\propto 1/\tau_j $ arises as a special case  for $a_{\tau}=2$ and $b_{\tau}=0$ which is invariant under rescaling. Furthermore. it offers an interesting physical interpretation as it favors ``slower'' varying representations (i.e. smaller $\tau$'s). 
In order to automatically determine the mean of the Gamma prior, we express  $b_{\tau}=\mu_{j} a_{\tau} $ where $\mu_j$ is a location parameter for which  an Exponential hyper-prior is used with a hyper-parameter $a_{\mu}$ i.e. $p(\mu_j) =\frac{1}{a_{\mu}}  e^{-\mu_j/a_{\mu}}$. Integrating out the $\mu_j$'s leads to following prior:
\be
\label{eq:m99}
p(\{\tau_j\}_{j=1}^k \mid k, ~a_{\tau},a_{\mu} ) = \prod_{j=1}^k     \frac{\Gamma(a_{\tau}+1) }{ \Gamma(a_{\tau}) }~\frac{ a_{\tau}^{a_{\tau}} }{\tau_j^{(a_{\tau}-1)} }~\frac{1}{a_{\mu}}~\frac{1}{(a_{\tau}\tau_j+a_{\mu}^{-1})^{(a_{\tau}+1)} }  
\ee


For the coefficients $a_j$ a multivariate normal prior was adopted:
\be
\label{eq:m6}
 \{{a}_j\}_{j=0}^k \mid k, \sigma_a^2 \sim N(\bs{0}, \sigma_a^2~\bs{I}_{k+1} )
\ee
where $\bs{I}_{k+1}$ is the $(k+1)\times(k+1)$ identity matrix. 
The hyper-parameter $\sigma^2$ which controls the spread of the prior is modeled by the standard inverse gamma distribution $ Inv-Gamma(a_0,b_0)$. It can readily be marginalized leading to the following prior for $a_j$'s:
\be
\label{eq:m6a}
p(\{{a}_j\}_{j=0}^k \mid k, ~a_0,b_0)=\frac{1}{(2\pi)^{(k+1)/2} } \frac{\Gamma(a_0+\frac{k+1}{2} ) }{ \left( b_0+\frac{1}{2} \sum_{j=0}^k a_j^2 \right)^{a_0+(k+1)/2} }
\ee


For the unknown kernel center locations $\nu_j$, a uniform prior in $[0,1]^M$ was used. Naturally if prior information is available about  subregions with significant fluctuations  this can be incorporated in the prior.


Based on the aforementioned equations, the complete prior model is given by:
\begin{eqnarray}
\label{eq:prior}
p(\bs{\theta} \mid s,~a_{\tau},a_{\mu},~a_0,b_0) & = & \frac{1}{(s+1)^{k+1}} \nonumber \\
 & \times  & \prod_{j=1}^k     \frac{\Gamma(a_{\tau}+1) }{ \Gamma(a_{\tau}) }~\frac{ a_{\tau}^{a_{\tau}} }{\tau_j^{(a_{\tau}-1)} }~\frac{1}{a_{\mu}}~\frac{1}{(a_{\tau}\tau_j+a_{\mu}^{-1})^{(a_{\tau}+1)} }  \nonumber \\
 & \times & \frac{1}{(2\pi)^{(k+1)/2} } \frac{\Gamma(a_0+\frac{k+1}{2} ) }{ \left( b_0+\frac{1}{2} \sum_{j=0}^k a_j^2 \right)^{a_0+(k+1)/2} }
\end{eqnarray}

Given $n$ data pairs, $(\bsf{x}_i,y_i)_{i=1}^n$ the likelihood $p(y_{1:n} \mid \bsf{x}_{1:n}, \bt)$ is: 
\begin{eqnarray}
\label{eq:like}
p(y_{1:n} \mid \bsf{x}_{1:n}, \bt) & = & \prod_{i=1}^n p(y_i \mid \bsf{x}_i, \bt) \nonumber \\
& = & \frac{1}{(2\pi)^{n/2}} \frac{1}{\sigma^n} \exp\{ -\frac{1}{2\sigma^2} \sum_{i=1}^n (y_i-f(\bsf{x}_i; \bt))^2 \}
\end{eqnarray}
A $Gamma(a,b)$ prior  was used for the variance $\sigma^{-2}$ of the Gaussian error in \refeq{eq:def3}, which is conjugate to the likelihood above and can be readily marginalized resulting to the following expression:
\be
\label{eq:like1}
L_n(\bt)=p(\bt  \mid (\bsf{x}_{1:n},y_{1:n} ) )=    \frac{\Gamma (a+n/2) }{ \left( b+\frac{1}{2}  \sum_{i=1}^n (y_i-f(\bsf{x}_i; \bt))^2 \right)^{a+n/2} }
\ee
where $\Gamma(z)=\int_0^{+\infty} t^{z-1} ~e^{-t}~dt$ is the gamma function.

The combination of the prior   $p(\bs{\theta})$ with the likelihood $L_n(\bs{\theta})$ corresponding to $n$ data points, give rise to the {\em posterior} density $\pi_n(\bs{\theta})$ which is proportional to:
\be
\label{eq:m13}
\pi_n(\bs{\theta})=p_n(\bs{\theta} \mid (\bsf{x}_{1:n},y_{1:n} ) ) \propto L_n(\bs{\theta}) ~p(\bs{\theta}) 
\ee

Even though several parameters have been marginalized  from the pertinent expressions, the corresponding posteriors can be readily be obtained, or rather be sampled from, once the posteriors $\pi_n(\bs{\theta})$ has been determined. Of particular interest for prediction purposes is the variance $\sigma^2$ of the error term (\refeq{eq:def3}).
 From \refeq{eq:like} and the conjugate prior model adopted for $\sigma^2$, it can readily be shown that the conditional posterior is given by a Gamma distribution:
\begin{eqnarray}                                                                              \label{eq:m12}
\pi_n(\sigma^{-2} ,\bs{\theta}) & = & p(\sigma^{-2} ,\bs{\theta} \mid  (\bsf{x}_{1:n},y_{1:n} )) \nonumber \\
 & = & \pi_n(\sigma^{-2} \mid \bs{\theta} ) ~ \pi_n(\bs{\theta} \mid  (\bsf{x}_{1:n},y_{1:n} ))
\end{eqnarray}
and:
\begin{eqnarray}
\label{eq:m12a}
\pi_n(\sigma^{-2} \mid \bs{\theta} )& = &p(\sigma^{-2} \mid \bs{\theta}, (\bsf{x}_{1:n},y_{1:n} ) ) \nonumber \\
&  = &  Gamma \left( a+\frac{n}{2}, b+ \frac{ \sum_{i=1}^n ( y_i-f(\bsf{x}_i; \bt) )^2 }{ 2 } \right)
\end{eqnarray}
In the context of Monte Carlo simulation, this trivially implies that once samples $\bs{\theta}$ from $\pi_n$ have been obtained, samples of $\sigma^{-2}$ can also be drawn from the aforementioned Gamma.

 It is worth pointing out, that \refeq{eq:m13} defines a {\em sequence of posterior densities} with support on $\cup_{k=0}^{k_{max}} \{k\} \times \bs{\Theta}_k$. Each $\pi_n$  corresponds to $n$ data points. It is easily understood that  for small datasets, i.e. small  $n$,  the effect of the likelihood function $L_n$ will be subdued and the the associated posterior $\pi_n$ will  have fewer modes as it is dominated by the prior. As more data points are added and $n$ increases the contribution of the likelihood becomes more pronounced and the posterior will potentially exhibit more idiosyncratic characteristics. As a result the task of  identifying these posteriors becomes increasingly more difficult for larger $n$. It is this feature that we propose of exploiting in the next section in order to increase the accuracy and improve on the efficiency of the inference process. 
  

\subsection{Bayesian Computation - Determining the Posterior }
\label{inference}

The posterior defined above is analytically intractable. For that reason, {\em Monte Carlo} methods provide essentially the only accurate way to infer $\pi_n$. Traditionally {\em Markov Chain Monte Carlo} techniques (MCMC) have been employed to carry out this task (\cite{tip01spa,and01rob}). These are based on  building a Markov chain that asymptotically converges to the target density (in this case $\pi_r$) by appropriately defining a transition kernel.  While convergence can be assured under weak conditions (\cite{liu01mon,rob04mon}), the rate of convergence can be extremely slow and require a lot of likelihood evaluations. Particularly in cases where the target posterior can have multiple modes, very large {\em mixing times} might be required.  In this work we propose a recursive inference algorithm based on {\em Sequential Monte Carlo} techniques (SMC,  \cite{mac98seq,dou01seq}) that ingests the data one at a time or in larger batches and independently of the order of presentation.  The sequential incorporation of data points introduces a tempering effect in the sense described in the previous paragraph. As a result, the global problem of identifying a potentially multi-modal posterior is decomposed to a series of easier, tractable problems. More importantly, the inferences made can be readily updated if more data becomes available.  
   As with Markov Chain Monte Carlo methods (MCMC), in SMC samplers the target
distribution(s) need only be known up to a constant and therefore
 do not require calculation of the intractable integral in the denominator in \refeq{eq:m0}.  
 The basis of the approximation is a  set of random samples (commonly referred to as {\em particles}), which  are propagated  using a combination of {\em importance sampling},  {\em resampling}  and MCMC-based {\em rejuvenation} mechanisms (\cite{mor06seq,del06seq}). Each of these particles is associated with an {\em importance  weight} which is proportional to the the posterior value of the respective particle. 
These weights are updated sequentially along with the particle locations. Hence if $\{\bs{\theta}_n^{(i)},~w_n^{(i)} \}_{i=1}^N$ represent $N$ such particles and associated weights for distribution $\pi_n(\bs{\theta})$ then:
\be
\label{eq:m14}
\pi_n(\bs{\theta}) \approx \sum_{i=1}^N ~W_n^{(i)}~ \delta_{\bs{\theta}_n^{(i)}} (\bs{\theta})
\ee
where $W_n^{(i)}=w_n^{(i)}/\sum_{i=1}^N w_n^{(i)}$ are the normalized weights and  $\delta_{\bs{\theta}^{(i)}_n }(.)$ is the Dirac function centered at $\bs{\theta}_n^{(i)}$. Furthermore, for any function $h(\bs{\theta})$  which is $\pi_n$-integrable (\cite{mor04fey,Chopin:2004}):
\be
\label{eq:m15}
 \sum_{i=1}^N ~W_n^{(i)}~h(\bs{\theta}_n^{(i)})  \rightarrow \int h(\bs{\theta})~\pi_n(\bs{\theta})~d\bs{\theta} \quad \textrm{almost surely}
\ee

In order to facilitate the transition between two successive posteriors $\pi_n$ and $\pi_{n+1}$ (particularly for small $n$), we can introduce a series of {\em bridging} distributions, based on a modified annealing scheme. In particular, if $\pi_0$ is the prior $p(\bt)$ (\refeq{eq:prior}) we define a family of artificial, auxiliary distributions $\pi_{n,\gamma}(\bt)$ as follows:
\be
\label{eq:m16}
\pi_{n,\gamma}(\bs{\theta}) \propto L_{n,\gamma}(\bt) ~p(\bs{\theta}) 
\ee
based on the modified likelihood:
\be
\label{eq:m16a}
L_{n,\gamma}(\bt)= \frac{\Gamma (a+(n+\gamma)/2) }{ \left( b+\frac{1}{2}  \sum_{i=1}^n (y_i-f(\bsf{x}_i; \bt))^2 +\gamma (y_{n+1}-f(\bsf{x}_{n+1}; \bt))^2 \right)^{a+(n+\gamma)/2} } \qquad \gamma \in [0,1]
\ee
where $\gamma$ plays the role of {\em reciprocal temperature}. Trivially for $\gamma=0$ we recover $\pi_n$ and for $\gamma=1$, $\pi_{n+1}$. The role of these auxiliary distributions is to {\em bridge the gap  between $\pi_n$ and $\pi_{n+1}$} and provide a smooth transition path where importance sampling can be efficiently applied. In this process, inferences based on $n$ data points  are  {\em transferred  and updated} to conform with additional $(n+1)^{th}$ datum. Starting with a particulate approximation for $\pi_0(\bs{\theta})=p(\bs{\theta})$ (which trivially involves  drawing samples from the prior with weights $w_0^{(i)}=1$), the goal is to gradually update the importance weights and particle locations in order to approximate the target posteriors $\pi_n$.

We propose an adaptive SMC algorithm, that extends existing versions (\cite{del06seq,mor06seq}) in that it automatically determines the number of intermediate bridging distributions needed.  In this process we are guided by the   Effective Sample Size $ESS=1/\sum_{i=1}^N (W_{s+1}^{(i)})^2$  which  provides a measure of degeneracy in the population of particles. Let $s$ denote the number of intermediate bridging distributions between $\pi_n$ and $\pi_{n+1}$ and $\gamma_s$ the associated reciprocal temperature. If $ESS_s$ is the $ESS$ of the population after the step $s$, then in the most favorable scenario that the next bridging distribution $\pi_{n,\gamma_{s+1}}$ is very similar to $\pi_{n,\gamma_{s}}$, then  $ESS_{s+1}$ should not be that much different from $ESS_s$. On the other hand if that difference is pronounced then $ESS_{s+1}$ could drop dramatically. Hence in determining, the next auxiliary distribution, we define an  acceptable reduction in the $ESS$, i.e. $ESS_{s+1} \ge \zeta ~ESS_s$ (where $\zeta<1$) and prescribe $\gamma_{s+1}$ (\refeq{eq:m16}) accordingly.
The proposed Adaptive SMC algorithm is summarized in Table \ref{tab:asmc}.

\begin{table}[htp!]
\caption{Basic steps of the {\em Adaptive } SMC algorithm proposed}
\begin{center}
\begin{tabular}{|p{12cm}|}
\hline
\noindent {\bf Adaptive SMC algorithm:}
\begin{enumerate}
\item Initialize population $\{\bs{\theta}_{0,0}^{(i)},~w_{0,0}^{(i)} \}_{i=1}^N$ where $\bs{\theta}_{0,0}^{(i)}$ are i.i.d draws from the prior $\pi_0$ and $w_{0,0}^{(i)}=1$ ($ESS_0=0$). Set $l=0$ and  $s=0$ and $\gamma_0=0$.
\item For $l < n$:
\bi
\item[a)] Set $s=s+1$.

\item[b)] {\em Reweigh}: If  $w_{l,s}^{(i)}(\gamma_s) =w_{l,s-1}^{(i)}~\frac{ \pi_{l,\gamma_{s}}(\bs{\theta}_{l,s-1}^{(i)}) }{ \pi_{l,\gamma_{s-1}}( \bs{\theta}_{l,s-1}^{(i)}) }$ are the {\em updated } weights as a function of $\gamma_s$ then determine $\gamma_s\in (\gamma_{s-1},~1]$ so that the associated $ESS_s = \zeta ESS_{s-1}$ (the value $\zeta=0.95$ was used in all the examples). Calculate $w_{l,s}^{(i)}$  for this $\gamma_s$.

\item[c)] {\em Resample}: If $ESS_s \le ESS_{min}$ then resample (the value $ESS_{min}=N/2$ was used in all the examples).

\item[d)] {\em Rejuvenate}: Use an MCMC kernel $P_{l,s}(.,.)$ that leaves $\pi_{l,\gamma_s}$
invariant to perturb each particle $\bs{\theta}_{l,s-1}^{(i)} \to \bs{\theta}_{l,s}^{(i)}$

\item[e)] The current population $\{\bs{\theta}_{l,s}^{(i)}, w_{l,s}^{(i)} \}_{i=1}^N$ provides a particulate approximation of $\pi_{l,\gamma_s}$  in the sense of Equations        (\ref{eq:m14}), (\ref{eq:m15}).

\item[f)] If  $\gamma_s =1$  set $l=l+1$, $\bs{\theta}_{l,0}^{(i)}=\bs{\theta}_{l-1,s}^{(i)}$, $w_{l,0}^{(i)}=w_{l-1,0}^{(i)}$, $s=0$ and  $\gamma_0=0$
\ei
\end{enumerate} \\
\hline
\end{tabular}
\end{center}
\label{tab:asmc}
\end{table}

 The role of the {\em Reweighing} step is to correct for  the discrepancy between the two successive distributions in exactly the same manner that importance sampling is employed. The {\em Resampling} step aims at reducing the variance of the particulate approximation by eliminating particles with small weights and multiplying the ones with larger weights.
The metric that we use in carrying out this task is the Effective Sample Size (ESS) defined earlier. If this degeneracy exceeds a specified threshold,  resampling is performed. As it has been pointed out in several studies (\cite{dou06eff}), frequent resampling can  deplete the population of its informational content and result in particulate approximations that consist of even a single particle. Throughout  this work $ESS_{min}=N/2$ was used.
 Although other options are available, {\em multinomial} resampling is most often applied and was found sufficient in the problems examined.

  A critical component  involves the perturbation of the population of samples by a standard MCMC kernel in the {\em Rejuvenation} step as this determines how fast the transition takes place. Although there is  freedom in selecting the transition kernel $P_s(.,.)$ (the only requirement is that it is $\pi_{l,\gamma_s}$-invariant), there is a  distinguishing feature that will be elaborated  in the next sub-section (see \ref{rjmcmc}).  The target posteriors $\pi_n$ (as well as the intermediate bridging distributions in \refeq{eq:m16}) live in spaces of varying dimensions as previously discussed. Hence an exploration of the state space must involve {\em trans-dimensional } proposals. Pairs of such moves can be defined in the context of Reversible-Jump MCMC (RJMCMC , \cite{Green:1995}) such as {\em adding/deleting} a kernel in the expansion of \refeq{eq:m2}, or {\em splitting/merging} kernels (see \ref{rjmcmc}). 


 It should  be noted that the framework proposed is directly {\em parallelizable}, as the evolution (reweighing, rejuvenation) of each particle is {\em independent} of the rest. 
 The particulate approximations obtained at each step, provide a {\em concise } summary of the posterior distribution based on the respective forward solver. This can be readily updated in the manner explained above, if more data become available, i.e. more runs of the approximate and exact solver are invoked. 

 An advantageous feature of the proposed framework is that the confidence in the estimates made can be readily quantified by establishing posterior (or credible) intervals from the particulate approximations (\refeq{eq:m14}). It is these credible intervals (or in general measures of the variability in the estimates such as the posterior variance) that can guide {\em adaptive acquisition of data}. Since we want to minimize calls to the exact solver $y$, we can utilize these inferences in order to perform runs in regions that will be  most informative of the sought  output and therefore make near-optimal use of the computational resources available. This will be discussed in more detail in the numerical examples.

\subsubsection{Trans-dimensional  MCMC}
\label{rjmcmc}

As mentioned earlier, a critical component in the SMC framework proposed is the MCMC-based rejuvenation step of the particle locations $\bs{\theta}$. It should be noted that the kernel $P_s(.,.)$ in the rejuvenation step (Step 3 of the SMC algorithm) need not be known explicitly as it does not enter in any of the pertinent equations. It is suffices that it is $\pi_{12,\gamma_s}$-invariant which is the target density. For the efficient exploration of the state space, we employ a mixture of moves which involve fixed dimension proposals (i.e. proposals for which the cardinality of the representation $k$ is unchanged) as well as moves which alter the dimension $k$ of the vector of parameters $\bs{\theta}$. We consider a total of $M=7$ such moves, each selected with a certain probability as discussed below. Of those, four involve trans-dimensional proposals which warrant a more detailed discussion. 

It is generally difficult to design proposals that alter the dimension significantly while ensuring a reasonable acceptance ratio. For that purpose, in this work we consider proposals that alter the cardinality  $k$ of the expansion by $1$ i.e. $k'=k-1$ or $k'=k+1$. We adopt the  the Reversible-Jump MCMC (RJMCMC) framework introduced in \cite{Green:1995} according to which such moves are defined in pairs in order to ensure reversibility of the Markov kernel (even though the reversibility  condition is not necessary, it greatly facilitates the formulations). We consider two such pairs of moves, namely {\em birth-death} and {\em split-merge}. Let  a proposal from $(k,\bs{\theta})$ to $(k',\bs{\theta'})$ that increases the dimension i.e.  $k='k+1$ and $\bs{\theta} \in \RR_{m_k}$, $\bs{\theta'} \in \RR^{m_{k+1}}$ (see last paragraph of sub-section \ref{prior}). Let $p(k \to k')$ the probability that such a proposal is made (user specified) and $p(k' \to k)$ the probability that the {\em reverse}, dimension-decreasing proposal is made. In order to account for the $m_{k+1}-m_k$ difference  in the dimensions of  $\bs{\theta}$ and $\bs{\theta'}$, the former is augmented with a vector $\bs{u} \in \RR ^{m_{k+1}-m_k}$ drawn from a distribution $q(\bs{u})$. Consider a differential and one-to-one mapping $h: \RR^{m_{k+1}} \to \RR^{m_{k+1}} $ that connects the three vectors as $\bs{\theta'}=h(\bs{\theta},\bs{u})$. Then as it is shown in \cite{Green:1995}, the acceptance ratio of such a proposal is:
\be
\label{eq:rjmcmc}
\min \left\{ 1, \frac{\pi_{12,\gamma_s}(\bs{\theta'}) p(k \to k') }{ \pi_{12,\gamma_s}(\bs{\theta})p(k' \to k) }  \frac{1}{q(\bs{u})} \left| \frac{\partial \bs{\theta'}}{ \partial (\bs{\theta},\bs{u} )} \right| \right\}
\ee
where  $ \left| \frac{\partial \bs{\theta'}}{ \partial (\bs{\theta},\bs{u}) } \right| $ is the Jacobian of the mapping $h$. Such a proposal is invariant w.r.t. the density $\pi_{12,\gamma_s}$. Similarly one can define, the acceptance ratio of the {\em reverse}, dimension-decreasing move:
\be
\label{eq:rjmcmc1}
\min \left\{ 1, \frac{\pi_{12,\gamma_s}(\bs{\theta}) p(k' \to k) }{ \pi_{12,\gamma_s}(\bs{\theta'})p(k \to k') }  q(\bs{u}) \left| \frac{\partial \bs{\theta'}}{ \partial (\bs{\theta},\bs{u}) } \right|^{-1} \right\}
\ee

In the following we provide details for the reversible pairs used in this work.

\noindent {\bf Birth-Death:} In order to simplify the resulting expressions, we assign the following  probabilities of proposing one of these moves $p_{birth}=c~min\{1, \frac{p(k+1)}{p(k)} \}=c~ \frac{1}{s+1}$ (from \refeq{eq:m5a}) and $p_{death}=c~min\{1, \frac{p(k-1)}{p(k)} \}=c $ (from \refeq{eq:m5a}). The constant  $c$  is user-specified (it is taken equal to $0.2$ in this work). Obviously if $k=k_{max}$, $p_{birth}=0$ and if $k=0$, $p_{death}=0$.

For the death move:
\bi 
\item A kernel $j$ ($1 \le j \le k$ ) is selected uniformly and removed from the representation in \refeq{eq:m2}.
\item The corresponding $a_j$ is also removed.
\ei
For the birth move:
\bi
\item A new kernel $k+1$ is added to the expansion while the existing terms remain unaltered.
\item The associated amplitude $a_{k+1}$ is drawn from $\mathcal{N}(0,\sigma_4^2)$ (the variance $\sigma_4^2$ is equal to the average of the squared amplitudes $a_j$ over  all the particles at the previous iteration)
\item The associated scale parameter $\tau_{k+1}$ is drawn from the prior, \refeq{eq:m99}
\item The associated kernel location $\bs{\nu}_{k+1}$ is also drawn from the uniform prior, \refeq{eq:prior}.
\ei
Hence the vector of dimension-matching parameters  $\bs{u}$ consists of $\bs{u}=(a_{k+1} ,$ $ \tau_{k+1}, \bs{x}_{k+1})$ and the corresponding proposal $q(\bs{u})$ is:
\be
\label{eq:rjmcmc2}
q(\bs{u})=\frac{1}{\sqrt{2\pi}} \frac{1}{\sigma_4}e^{-\frac{1}{2}~a_{k+1}^2/\sigma_4^2}~\frac{b_{\tau}^{a_{\tau}} }{\Gamma(a_{\tau}) } \tau_{k+1}^{ a_{\tau}-1 } ~\exp(-b_{\tau} \tau_{k+1})
\ee
It is obvious that the Jacobian of such a transformation is $1$. 

\noindent {\bf Split-Merge} These moves correspond to splitting an existing kernel into two  or merging two existing kernels into one. Similarly to the birth-death pair, they alter the dimension of the expansion by $1$ and are selected with probabilities  $p_{split}= \frac{1}{s+1}$ and $p_{merge}=c $. For obvious reasons, $p_{split}=0$ if $k=k_{max}$ and $p_{merge}=0$ if $k \le 1$.    Consider first the  merge move between two kernels $j_1$ and $j_2$. 
 In order to ensure a reasonable acceptance ratio, merge moves are only permitted when the (normalized) distance between  the kernels is relatively small and when  the  amplitudes $a_{j_1}$, $a_{j_2}$  are relatively similar. Specifically we require that the following two conditions are met: 
\be
\label{eq:rjmcmc3}
\frac{\parallel \bs{\nu}_{j_1}-\bs{\nu}_{j_2} \parallel }{ \sqrt{ \tau_{j_1}^{-1}+\tau_{j_2}^{-1} } } \le \delta_x \qquad \mid a_{j_1}-a_{j_2} \mid \le \delta_a
\ee
(the values $\delta_x =\delta_a=1$ were used in this work). Two candidate kernels are selected {uniformly} from the pool of pairs satisfying the aforementioned conditions. The proposed kernels  $j_1$ and $j_2$ are removed from the expansion and are  substituted by a new kernel $j$ with the following associated parameters:
\bi
\item \be
\label{eq:rjmcmc4}
\tau_{j}= \left(\sqrt{ \tau_{j_1}^{-1}+\tau_{j_2}^{-1} } \right)^{-1}
\ee
\item \be
\label{eq:rjmcmc5}
a_{j}=\sqrt{\tau_{j} }(\frac{a_{j_1}}{\sqrt{\tau_{j_1}} } + \frac{a_{j_2}}{\sqrt{\tau_{j_2}} } )
\ee
This ensures that the {\em average} value of the previous expansion (with $j_1$ and $j_2$) in \refeq{eq:m2} when integrated in $\RR^d$  is the same with the new (which contains $j$ in place of  $j_1$ and $j_2$)
\item \be
\label{eq:rjmcmc6}
\bs{\nu}_{j}=\frac{ \bs{\nu}_{j_1}+\bs{\nu}_{j_2} }{2}
\ee
\ei 

The split move is applied to a kernel $j$ (selected {\em uniformly}) which is substituted by two new kernels $j_1$, $j_2$.    In order to ensure {\em reversibility},  kernels $j_1$ and $j_2$ should satisfy the requirements of \refeq{eq:rjmcmc3} and the application of a merge move in the manner described above, should return to the original kernel $j$.  There are several ways to achieve this, corresponding essentially to different vectors $\bs{u}$ and mappings $h$ in \refeq{eq:rjmcmc}. In this work:
\bi
\item A scalar $u_{\tau}$ is drawn from the uniform distribution $U[0,1]$ and $\tau_{j_1}^{-1}=u_{\tau} \tau_j^{-1}$ and $\tau_{j_2}^{-1}=(1-u_{\tau}) \tau_j^{-1}$. This ensures compatibility with \refeq{eq:rjmcmc4}.
\item A vector $\bs{u}_x$ is drawn uniformly in the ball of radius $R$ where $R=\frac{\delta_x}{2\sqrt{\tau_j} }$. The center of the new kernels are specified as $\bs{\nu}_{j_1}=\bs{\nu}_j-\bs{u}_x$ and $\bs{\nu}_{j_2}=\bs{\nu}_j+\bs{u}_x$. This ensures compatibility with the first of \refeq{eq:rjmcmc3} as well as \refeq{eq:rjmcmc6}.
\item A scalar $u_a$ is drawn from the uniform distribution $U[-\frac{\delta_a}{2}, \frac{\delta_a}{2}]$. The amplitudes of the new kernels are determined by $a_{j_1}=\hat{a}-u_a$ and $a_{j_2}=\hat{a}+u_a$, where $\hat{a}=\frac{a+u_a(\sqrt{u_{\tau}}-\sqrt{1-u_{\tau} } ) }{ \sqrt{u_{\tau}}+\sqrt{1-u_{\tau} } }$. This ensures compatibility with the second of \refeq{eq:rjmcmc3} as well as \refeq{eq:rjmcmc5}.

\ei

The vector of dimension-matching parameters  $\bs{u}$ (in \refeq{eq:rjmcmc}) consists of $\bs{u}=(u_{\tau}, \bs{u}_x, u_a)$ and the corresponding proposal $q(\bs{u})$ is a product of uniforms in the domains specified above.
After some algebra, it can be shown that the Jacobian of such a transformation is $2^{M+1} \frac{\tau}{ u_{\tau}^2 (1-u_{\tau})^2 } \frac{1}{  \sqrt{u_{\tau}}+\sqrt{1-u_{\tau} } }$. 

\begin{figure}[]
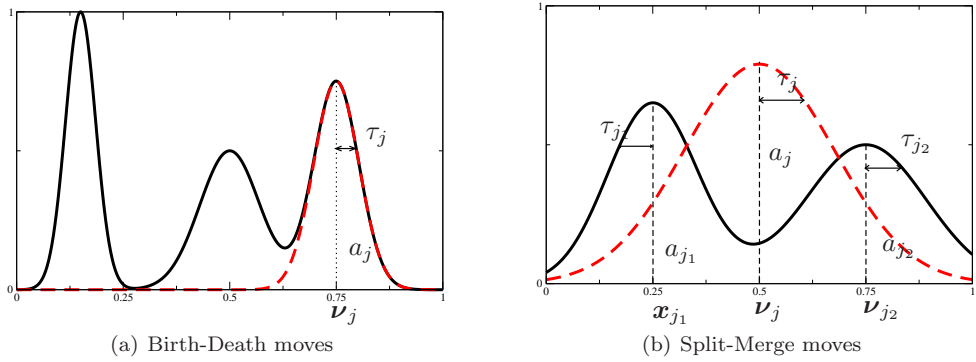

     \centering
     \subfigure[Birth-Death moves]{
          \label{fig:rjmcmc1}
\psfrag{a}{$a_j$}
\psfrag{t}{$\tau_j$}
\psfrag{x}{$\bs{\nu}_j$}
           \includegraphics[width=.45\textwidth]{FIGURES/rjmcmc1.eps}} \hfill
     \subfigure[Split-Merge moves]{
          \label{fig:rjmcmc2}
\psfrag{a}{$a_j$}
\psfrag{t}{$\tau_j$}
\psfrag{x}{$\bs{\nu}_j$}
\psfrag{a1}{$a_{j_1}$}
\psfrag{t1}{$\tau_{j_1}$}
\psfrag{x1}{$\bs{x}_{j_1}$}
\psfrag{a2}{$a_{j_2}$}
\psfrag{t2}{$\tau_{j_2}$}
\psfrag{x2}{$\bs{\nu}_{j_2}$}
             \includegraphics[width=.45\textwidth]{FIGURES/rjmcmc2.eps}}\\
     \label{fig:rjmcmc}
\caption{Trans-dimensional RJMCMC proposals}
\vspace{.5cm}
\end{figure}

The remaining three proposals, involve fixed-dimension moves that do not change the cardinality of the expansion but rather perturb  some of the terms involved. In particular, we considered updates of the amplitude $a_j$, scale $\tau_j$ or location $\bs{x}_j$ of a kernel $j$ selected {\em uniformly} (naturally, in the case of the amplitudes, the constant $a_0$ (\refeq{eq:m2}) is also a candidate for updating). Each of these three moves is proposed with probability $\frac{1}{3}(p_{birth}+p_{death}+p_{split}+p_{merge})=\frac{2~c}{3}(\frac{1}{s+1}+1)$. In particular:
\begin{enumerate} 
\item Update $a_j \to a'_j$: A coefficient $a_j$ (in \refeq{eq:m2}) is {\em uniformly} selected and perturbed as: 
\be
\label{eq:mcmc1}
a'_j =a_j + \sigma_1~Z \quad, Z\sim\mathcal{N}(0,1)
\ee
\item Update $\tau_j \to \tau'_j$: A scale parameter $\tau_j$ (in \refeq{eq:m2}) is {\em uniformly} selected and perturbed as: 
\be
\label{eq:mcmc2} 
\tau'_j =\tau_j e^{\sigma_2 Z}, \quad Z\sim\mathcal{N}(0,1)
\ee
(this ensures positivity of $\tau'_j$)
\item Update $\bs{\nu}_j \to \bs{\nu'}_j$: A location $\bs{\nu}_j \in [0,1]^M $ (in \refeq{eq:m2}) is {\em uniformly} selected and perturbed as:
\be
\label{eq:mcmc3} 
\bs{\nu'}_j =\bs{\nu}_j+\sigma_3 ~\bs{Z}, \quad \bs{Z}=(Z_1,\ldots,Z_d), ~ Z_i \sim \mathcal{N}(0,1)
\ee
\end{enumerate}
The acceptance ratios are calculated based on the standard MCMC formulas using $\pi_{{12},\gamma_s}$ as the target density. 
It should be noted that the variances in the random walk proposals are adaptively selected so that the respective acceptance rates are in the range $0.2 - 0.4$. As it is well-known  (chapter 7.6.3 in \cite{rob04mon}) adaptive adjustments of Markov Chains based on  past samples can breakdown ergodic properties and lead to convergence issues in standard MCMC contexts. In the proposed SMC framework however, such restrictions do not apply as it suffices that the MCMC kernel is invariant. This is an additional advantage of the proposed simulation scheme in comparison to traditional MCMC.

\subsection{Prediction - Calculating statistics of exact solver}
\label{prediction}

We return to the original problem of estimating statistics of the output $y$ of the exact solver using the relationship with the outputs $\bs{x}$ of the approximate solvers. The aforementioned Bayesian model is able to not only  provide estimates but also  quantify the level of confidence one can assign to the predicted outcome. 
Let $({\bt},~{\sigma})$ denote a pair of values for the model parameters in \refeq{eq:def3}. 
For these parameter values,  the conditional density $p(y \mid \bsf{x})$ can be obtained based on the regression model adopted and \refeq{eq:def4}:
\be
\label{eq:def4a}
p(y \mid \bsf{x}) \approx p(y \mid \bsf{x},  {\bt}, {\sigma})=\frac{1}{\sqrt{2\pi}} \frac{1}{{\sigma}} \exp \{-\frac{1}{2{\sigma}^2} (y~-~ f(\bsf{x}; {\bt}))^2 \}
\ee
which upon substitution in \refeq{eq:def1a} yields:
\be
\label{eq:def1b}
Pr [ y  \in \mathcal{A} ;  {\bt}, {\sigma}] = \int q_{\mathcal{A}}(\bs{x}; {\bt}, {\sigma})~\pi_x(\bsf{x})~ d\bsf{x}
\ee
where:
\be
\label{eq:def1c}
q_{\mathcal{A}}(\bs{x}; {\bt}, {\sigma})=\int  \bs{1}_{\mathcal{A}}(y)~ p(y \mid \bsf{x},  {\bt}, {\sigma} ) ~dy 
\ee
The latter expresses the probability that exact response $y \in \mathcal{A}$ for fixed approximate response $\bsf{x}$ (and model parameters).
Consider for example the case that we are interested in calculating a probability of  exceeding a threshold $y_0 \in \RR$, i.e. $\mathcal{A}=(y_0, ~+\infty)$. Then:
\be
\label{eq:def1d}
q_{\mathcal{A}}(\bs{x}; {\bt}, {\sigma})= \Phi \left( \frac{f(\bsf{x}; {\bt}) -y_0}{{\sigma}} \right) 
\ee
where  $\Phi(z)=\int_{-\infty}^{z} \frac{1}{\sqrt{2\pi}} e^{-\frac{w^2}{2} }~dw$ is the standard normal CDF.

Given a number of training data $(\bsf{x}_{1:n}, ~y_{1:n})$,  the plausibility of various parameter values is quantified by the posterior $\pi_n(\bt)=p(\bt \mid  (\bsf{x}_{1:n}, ~y_{1:n}))$. Hence, one can obtain point estimates of $q_{\mathcal{A}}$ based for example on the maximum a posteriori values (MAP) $\bt_{MAP}=argmax~ \pi_n(\bt)$  or the posterior mean $E_n[\bt]=\int \bt ~\pi_n(\bt)~d\bt$. More importantly, due to its dependence of $\bt$ (and $\sigma$), $q_A$ is also random and its distribution can be determined from the posterior distribution of $\bt$ and $\sigma$. This distribution therefore indirectly depends on the training data upon which  posterior inferences were based. In particular, one can estimate the {\em posterior mean} of $q_{\mathcal{A}}$ for the case of \refeq{eq:def1d} as follows:
\begin{eqnarray}
\label{eq:def1e}
\hat{q}_{\mathcal{A}}(\bsf{x})=E_n[q_{\mathcal{A}}] & = & \int \Phi \left( \frac{f(\bsf{x}; {\bt}) -y_0}{{\sigma}} \right) ~\pi_n(\bt, \sigma^{-2})~d\bt~d\sigma^{-2} \nonumber \\
& = & \int \Phi \left( \frac{f(\bsf{x}; {\bt}) -y_0}{{\sigma}} \right) ~\pi_n(\bt)~\pi_n(\sigma^{-2} \mid \bt) ~d\bt~d\sigma^{-2} \nonumber \\
 & & \qquad \qquad \qquad \qquad \qquad \qquad (\textrm{from \refeq{eq:m12} }) \nonumber \\
& \approx & \sum_{i=1}^N W_n^{(i)}~ \Phi \left( \frac{f(\bsf{x}; \bt_n^{(i)}) -y_0}{{\sigma^{(i)}_n }} \right)
\end{eqnarray}
 where the particulate approximation $\{ \bs{\theta}_n^{(i)}, W_n^{(i)} \}_{i=1}^N$ of $\pi_n$ obtained through the proposed SMC scheme was used and $\{ \sigma_n^{(i)} \}_{i=1}^N$ are corresponding draws from the conditional posterior in \refeq{eq:m12a}.
\refeq{eq:def1e} expresses how, {\em on average, based on the available data $(\bsf{x}_{1:n}, ~y_{1:n})$} the probability of interest depends on the approximate solver values $\bsf{x}$. This can provide an estimate of  the probability of interest by substitution in \refeq{eq:def1b}.

Furthermore, the weighted samples $q_{\mathcal{A}}^{(i)}(\bsf{x})=\Phi \left( \frac{f(\bsf{x}; \bt_n^{(i)}) -y_0}{{\sigma^{(i)}_n }} \right)$ provide a particulate approximation of the distribution of $q_A(\bsf{x}),~\forall \bsf{x}$. p-Quantiles $q_{\mathcal{A},p}(\bsf{x})$ can be readily estimated as:
\be
\label{eq:def5}
Pr[q_A(\bsf{x}) \le q_{\mathcal{A},p}(\bsf{x})]\approx \sum_{i=1}^N  W_n^{(i)}~ H \left( q_{\mathcal{A}}^{(i)}(\bsf{x})-q_{\mathcal{A},p}(\bsf{x}) \right)=p
\ee
where $H(.)$ is the Heaviside function. These can readily yield {\em confidence bounds} for the probability of interest  when substituted in \refeq{eq:def1b}. More importantly perhaps, these bounds can serve as the basis for {\em active learning} i.e.  determining where more training samples need to be generate in order to refine the estimates produced.
Consider for example the $p=1\%$ and $p=99\%$ quantiles  $q_{\mathcal{A},0.01}(\bsf{x})$ and $ q_{\mathcal{A},0.99}(\bsf{x})$. then based on \refeq{eq:def1b}) an ordering of $\bs{x}$ can be constructed based on:
\be
\label{eq:def6}
\left( q_{\mathcal{A},0.99}(\bsf{x})-q_{\mathcal{A},0.01}(\bsf{x}) \right)~\pi_x(\bsf{x})
\ee
Thus $\bsf{x}$ (or regions in the $\bsf{x}$-space) for which the aforementioned value is large contribute more in the uncertainty about the probability of interest $Pr[y \in \mathcal{A}]$ and therefore could serve as the best candidates for generating additional training pairs $(\bsf{x}_{n+1}, y_{n+1})$. This is particularly important in the cases considered where each run for the evaluation of the exact response ${y}$ can be extremely expensive and therefore optimal use of the computational resources is crucial.
These additional training samples can be readily incorporated based on the SMC scheme adopted and updates of the particulate approximation that reflect the new data can be produced. These in turn can lead to updates in the estimates made as well as the confidence bounds.
Naturally other measures of variability of $q_{\mathcal{A}}(\bsf{x})$, such as  the  variance (or coefficient of variation) can be used in place of $ \left( q_{\mathcal{A},0.99}(\bsf{x})-q_{\mathcal{A},0.01}(\bsf{x}) \right)$ in \refeq{eq:def6}. The variance for each $\bsf{x}$ can be  estimated as follows:
\be
\label{eq:def1f}
Var_n[q_{\mathcal{A}}]  \approx  \sum_{i=1}^N W_n^{(i)}~ \left(  q_{\mathcal{A}}^{(i)}(\bsf{x}) -  \hat{q}_{\mathcal{A}}(\bsf{x})\right)^2
\ee

For estimates of expectations of functions $h(.)$ of the exact output $y$ as in \refeq{eq:def2a}, the same results apply if in place of $q_{\mathcal{A}}(\bsf{x})$  we use:
\be
\label{eq:def2b}
q_h(\bsf{x})=\int h(y) p(y\mid \bsf{x}) ~dy
\ee

\section{Numerical results}
\label{examples}
In the examples presented, the following values for the hyperparameters of the prior model were used (\refeq{eq:prior}):
\bi
\item $k_{max}=100$ and $s=1.0$ (\refeq{eq:m5a})
\item $a_{tau}=1.0$ (\refeq{eq:m9}) and $a_{\mu}=0.01$ (\refeq{eq:m99})
\item $a_0=1.0$ and $b_0=1.0$ (\refeq{eq:m6a})
\item $a=2.$ and $b=1. \times 10^{-6}$ (\refeq{eq:like})
\ei
Furthermore,  $N=1,000$ particles were employed in the adaptive  SMC scheme described in section \ref{inference}. As in most systems of practical interest, the computational cost  is dominated by the number of calls to the forward solver, we report results on  computational effort in terms of the number of runs of the {\em exact} solver.

\subsection{Example 1}

\begin{figure}
\psfrag{Tc}{$T_c$}
\psfrag{dc}{$\delta_c$}
\psfrag{Gc}{$G_c$}
\includegraphics[height=4.5cm,width=0.8\textwidth]{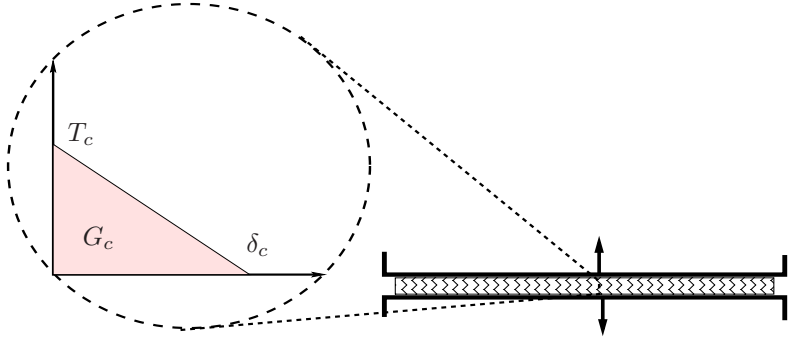}
\label{fig:ex21}
\caption{Cohesive interface - Cohesive Law: $T_c$ denotes ultimate interfacial tension (when the stress reaches $T_c$ the cohesive element is activated), $\delta_c$ denotes the ultimate separation interface (when the separation reaches $\delta_c$ the interface tension becomes zero) and $G_c$ denotes the fracture energy which is equal to the area under the tension-separation curve.}
\end{figure}

The first example involves  a problem from fracture mechanics where it is known that small scale stochastic fluctuations can have a significant impact in the macroscale response. We consider cohesive interface  of unit length that is pulled apart in Mode I fracture and is modeled  with cohesive zone elements (Figure \ref{fig:ex21}).  These are line (or surface in 3D) elements which are located at the interface and govern the separation process in accordance with a cohesive law. The concept of cohesive laws  was pioneered by Dugdale (\cite{dug60yie}) and Barenblatt (\cite{bar62mat}) in order to model fracture processes and  has
been successfully used in a Finite Element setting by several researchers (\cite{xu94num,cam96com,ort99fin}). According to these models, fracture is initiated when the interface traction exceeds a threshold $T_c$ and progresses gradually as the  separation takes place across an extended crack tip or cohesive zone and is resisted by
cohesive tractions. We assume herein a simple constitutive law relating interface traction-separation as seen in Figure \ref{fig:ex21}. Under monotonic
loading the normal interface traction decays as $T=T_c\left( 1-\frac{\delta}{\delta_c}\right)$ for $\delta \le \delta_c$ and $T=0$ for $\delta > \delta_c$. The fracture energy $G_c$ is given by $G_c=T_c \delta_c /2$. The constitutive rate equations are:
\be
\label{eq:coh}
\dot{T}=\left\{ \begin{array}{cc}
                -T_c \frac{\dot{\delta}}{\delta_c} & \textrm{if } \dot{\delta}>0 \\
                \frac{T}{\delta}~\dot{\delta} & \textrm{if } \dot{\delta} <0 \\                0 & \textrm{if } \delta > \delta_c \end{array} \right.
\ee
At the microstructural level, the cohesive properties exhibit random variability. 
 We adopt the following simple  random field descriptions for the  model parameters: 
\begin{eqnarray}
\label{eq:num4}
T_c(z) & = & T_0+ \Delta T_0 ~ U_1(z) \nonumber \\
G_c(z) & = & G_0+\Delta G_0 ~ (\rho U_1(z)+U_2(z)) \quad z \in [0,1]
\end{eqnarray}
where:
\be
\label{eq:num5}
U_i(z)= 2 \Phi (h_i(z))-1, ~i=1,2
\ee
and $h_1(z)$ is a zero-mean, unit variance Gaussian process with autocorrelation $R_h(\Delta z)=E\left[ h(z) h(z+\Delta z) \right]=exp \{-\frac{ \mid \Delta z \mid }{ z_0} \}$ ($\Phi$ is the standard normal CDF). The parameter $z_0$ controls the length scale of heterogeneity and it was taken equal to $0.1$. The field $h_2(z)$  was assumed to represent a Discretized white noise process. The parameter $\rho$ controls the autocorrelation between the two properties and was taken equal to $0.9$ which implies that areas with high $T_c$ are more likely to have high $G_c$ as well. Also, the values  $T_0=1.0$, $\Delta T_0=0.5$, $G_c=10^{-3}$ and $\Delta G_0=0.5 \times 10^{-3}$ were used.

The exact solver was a detailed  finite element model consisting of $1,000$  cohesive elements of equal length with properties assigned based on  the values of $T_c(z)$ and $G_c(z)$ at their midpoint. The mesh size is much smaller than the length scale of variability of the cohesive properties as determined by  the correlation length $z_0$ defined above. The output of interest $y$ was the fracture energy released when a uniform separation $\delta=0.5 \times 10^{-3}$ was applied at the interface. The quasi-static, nonlinear calculation of the output of interest was  carried out by applying separation increments of $0.5 \times 10^{-6}$ in order to capture accurately the traction-separation history (i.e. a total of $1,000$ iterations).

\begin{figure}[htp]
\vspace{.5cm}
\psfrag{exact}{exact $y$}
\psfrag{approx}{approximate $\bsf{x}$}
\psfrag{yx}{$y=\bsf{x}$}
\includegraphics[width=.8\textwidth,height=5.5cm]{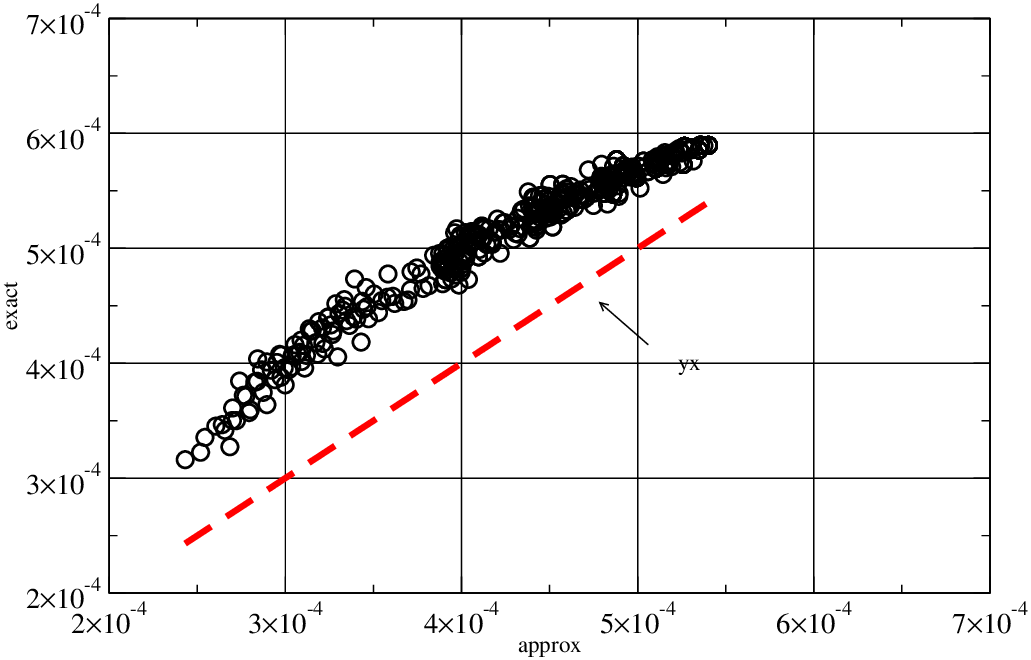}
\caption{Exact $y$ vs. Approximate $\bsf{x}$ response}
\label{fig:ex22}
\end{figure}

We considered  a single approximate solver with a much coarser mesh consisting of only $10$ cohesive elements of equal length, i.e. each macro-element takes the place of $100$ micro-elements of the exact solver. The assigned cohesive strength $T_c$ in each macro-element was set equal to the minimum of the cohesive strengths of the $100$ corresponding micro-elements,   and the fracture energy $G_c$ equal to the average of the fracture energies of the the $100$ corresponding micro-elements. In addition a  displacement increment of $ 0.5 \times 10^{-5}$ (in contrast to the $0.5 \times 10^{-6}$ for the exact solver) was used in order to carry out the quasi-static, nonlinear integration of the equations of equilibrium and the constitutive model (\refeq{eq:coh}).
As a result of these crude simplifications the approximate solver was $1,069$ faster than the exact. Figure \ref{fig:ex22} compares the approximate and exact solver output prediction where significant discrepancies can be observed (e.g. when $\bsf{x}= 4\times 10^{-4}$, $y \approx 5 \times 10^{-4}$ , i.e. a $25 \%$ difference). It is also observed that the mapping from $\bsf{x}$ to $y$  is one-to-many, as the coarse model crudely smears some of the fine details that affect the response.

In order to assess the performance of the method proposed we considered the event  $y > y_0=5.601 \times 10^{-4}$ which corresponds to a probability $10^{-3}$. This was found using an advanced simulation procedure  based on Sequential Monte Carlo as in the first example (\cite{au01est,kou08des}).  The computational effort amounted to  $1,500$ calls to the {\em exact} solver and the coefficient of variation (c.o.v) of the estimate was $0.23$ (based on an asymptotic, and rather optimistic, bound in \cite{au01est}). This roughly implies that with probability $0.95$, the actual probability of the event of interest is in the interval $[0.55 \times 10^{-3},~ 1.5 \times 10^{-3}]$. For the same c.o.v and standard Monte Carlo $18,500$ calls to the {\em exact} solver would have been needed.

\begin{figure}
\vspace{.5cm}
\psfrag{x0}{$x_0$}
\psfrag{px}{$Pr[x \ge x_0]$}
\includegraphics[width=.8\textwidth,height=5.5cm]{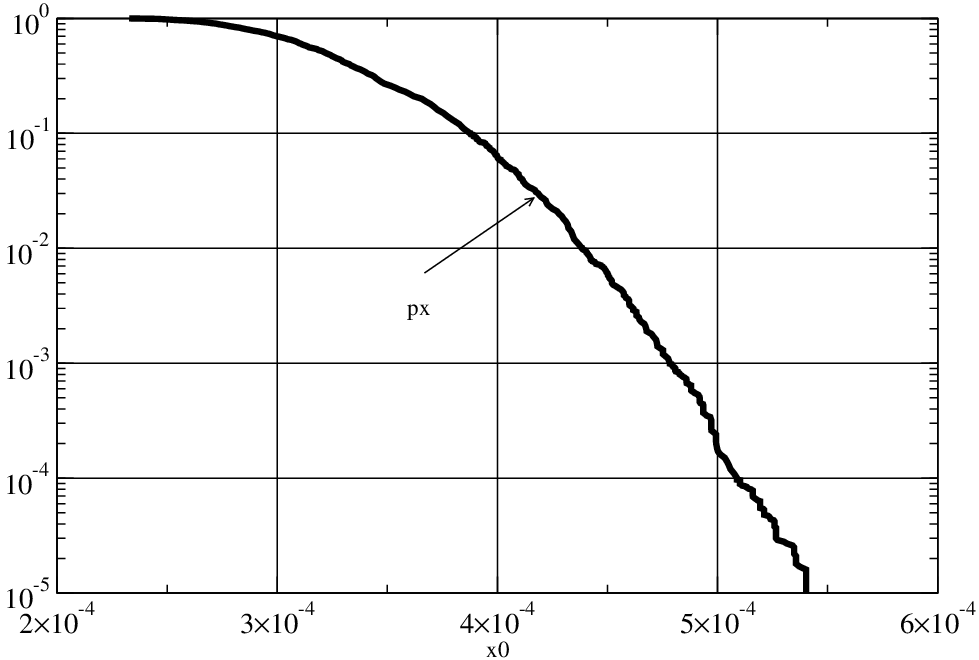}
\caption{Cumulative distribution function for the {\em approximate} solver output $\bs{x}$}
\label{fig:ex24}
\end{figure}

The density $\pi_{x}$ of $\bsf{x}=x_1$ was estimated using the same advanced Monte Carlo scheme that required $5,000$ calls to the {\em approximate} solver. The corresponding CDF is depicted in Figure \ref{fig:ex24} for probabilities as low as $10^{-5}$. Note that due to the reduced computational effort associated with calls to the approximate solver, the computational time for this task amounted to (approximately) $5$ runs of the {\em exact} solver.

The crucial task, that of estimating $p(y \mid x_1)$ involves the nonparametric Bayesian regression model discussed previously.  Figure \ref{fig:ex25} depicts posterior statistics of the regression model for various training sample sizes and Figure \ref{fig:ex26}  the posterior mean and posterior quantiles of $q_{\mathcal{A}}(\bsf{x})$ based on Equations (\ref{eq:def1e}) and \ref{eq:def5} for various sample sizes. Table \ref{tab:ex21} summarizes the estimates based on the posterior mean $\hat{q}_{\mathcal{A}}(\bsf{x})$  and confidence bounds established with  $q_{\mathcal{A},0.01}(\bsf{x})$ and $ q_{\mathcal{A},0.99}(\bsf{x})$.
It is noted that even with a small number of calls to the exact solver, the estimates obtained are reasonably good and most importantly the lower and upper confidence bounds always include the reference value. This is particularly important in engineering purposes as the analyst can decide whether these confidence bounds are satisfactory and if not perform additional calls to the exact solver in order to refine them. 
As the number of training samples increases the posterior mean approaches the true value and the credible intervals become more concentrated.
A complete view of the the cdf of the exact output $y$ is depicted in Figure 3.6 
 based on $150$ training samples.


%

\begin{figure}[htp]
    \centering
 \psfrag{approx}{approximate $x_1$}
  \psfrag{exact}{exact $y$}
     \subfigure[$10$ training samples ]{
          \label{fig:ex25a}
           \includegraphics[width=.48\textwidth,height=4.cm]{FIGURES/cohesive_samples_0010.eps}} \hfill
     \subfigure[$10$ training samples ]{
          \label{fig:ex25b}
\psfrag{sigma}{$\sigma$}
             \includegraphics[width=.48\textwidth,height=4.cm]{FIGURES/cohesive_posterior_error_0010.eps}}\\
\vspace{.2cm}
     \subfigure[$50$ training samples]{
          \label{fig:ex25e}
\psfrag{approx}{approximate $x_1$}
  \psfrag{exact}{exact $y$}
           \includegraphics[width=.48\textwidth,height=4.cm]{FIGURES/cohesive_samples_0050.eps}} \hfill
     \subfigure[$50$ training samples ]{
          \label{fig:ex25f}
\psfrag{sigma}{$\sigma$}
             \includegraphics[width=.48\textwidth,height=4.cm]{FIGURES/cohesive_posterior_error_0050.eps}}\\
\vspace{.2cm}
     \subfigure[$150$ training samples]{
          \label{fig:ex25g}
           \includegraphics[width=.48\textwidth,height=4.cm]{FIGURES/cohesive_samples_0150.eps}} \hfill
     \subfigure[$150$ training samples ]{
          \label{fig:ex25h}
\psfrag{sigma}{$\sigma$}
             \includegraphics[width=.48\textwidth,height=4.cm]{FIGURES/cohesive_posterior_error_0150.eps}}\\
     \caption{Posterior mean of $f(\bsf{x}; \bt)$ and and posterior density  of $\sigma$ (\refeq{eq:def3})  for various training sample sizes}
     \label{fig:ex25}
\end{figure}

\begin{figure}[htp]
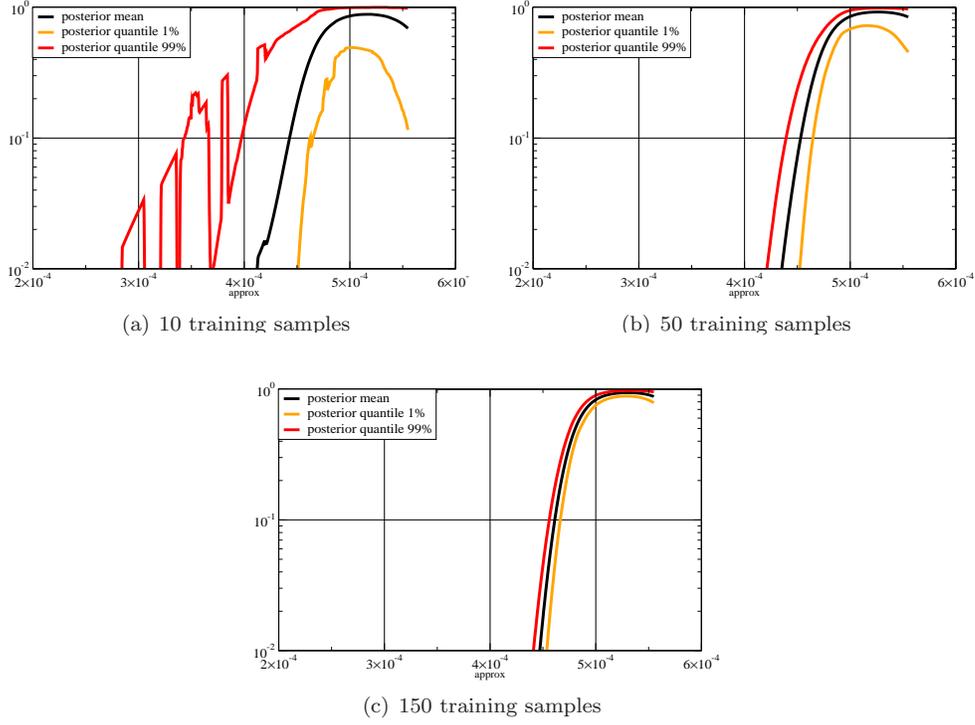

     \centering
     \subfigure[$10$ training samples]{
          \label{fig:ex26a}
           \includegraphics[width=.48\textwidth,height=4.cm]{FIGURES/cohesive_posterior_0010.eps}} \hfill
     \subfigure[$50$ training samples]{
          \label{fig:ex26b}
\psfrag{sigma}{$\sigma$}
             \includegraphics[width=.48\textwidth,height=4.cm]{FIGURES/cohesive_posterior_0050.eps}}\\
\vspace{.2cm}
     \subfigure[$150$ training samples]{
          \label{fig:ex26c}
           \includegraphics[width=.48\textwidth,height=4.cm]{FIGURES/cohesive_posterior_0150.eps}} \hfill
%
%
     \caption{Posterior mean and quantiles for $Pr[ y \ge 5.601 \times 10^{-4} \mid x]$ based on various sample sizes}
     \label{fig:ex26}
\end{figure}

\begin{table}[htp]
\caption{Estimates of $Pr[ y > y_0=5.601 \times 10^{-4}]$ and computational effort (the latter is measured in number equivalent number of calls to the {\em exact} solver}
\begin{center}
\begin{tabular}{c|c|c|c|c|}
\hline
Number & Posterior & Posterior & Posterior & Computational \\
of samples & mean  & quantile $1\%$ & quantile $99\%$ & Effort\\
\hline \hline
 10 &  $6.36\times 10^{-3}$ &   $5.91 \times 10^{-4}$ & $7.17 \times 10^{-2}$ &  $15$  \\
\hline
50 &  $ 1.75   \times 10^{-3}$  &  $7.39 \times 10^{-4}$   & $ 3.55 \times 10^{-3}$ & $55$ \\
\hline
 150 &  $1.01 \times 10^{-3}$ &   $7.07 \times 10^{-4}$ & $1.42 \times 10^{-3}$  & $155$ \\
\hline
\end{tabular}
\end{center}
\label{tab:ex21}
\end{table}

\begin{figure}
\vspace{.5cm}
\psfrag{y0}{$y_0$}
\psfrag{py}{$Pr[ y \ge y_0]$}
\includegraphics[height=4.5cm,width=0.8\textwidth]{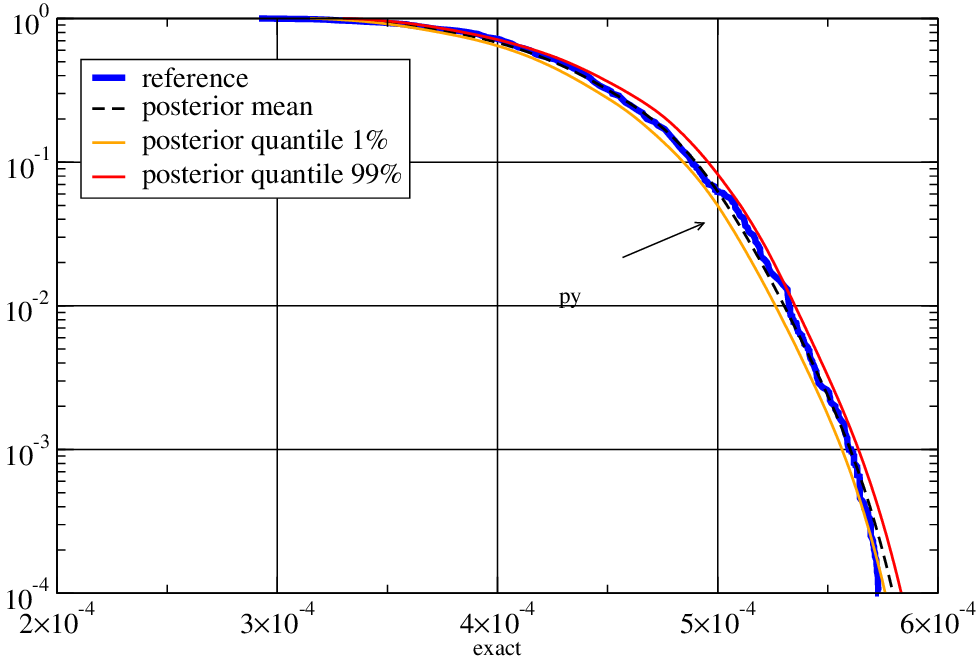}
\caption{Posterior mean and quantiles for $Pr[ y \ge y_0 ], ~ \forall y_0$ based on 150 training samples}
\label{fig:ex27}
\end{figure}


\subsection{Example 2}

\begin{figure}[htp]
     \centering
     \subfigure[Connected inclusion phase (high strength)]{
          \label{fig:ex11a}
           \includegraphics[width=.48\textwidth,height=5.5cm]{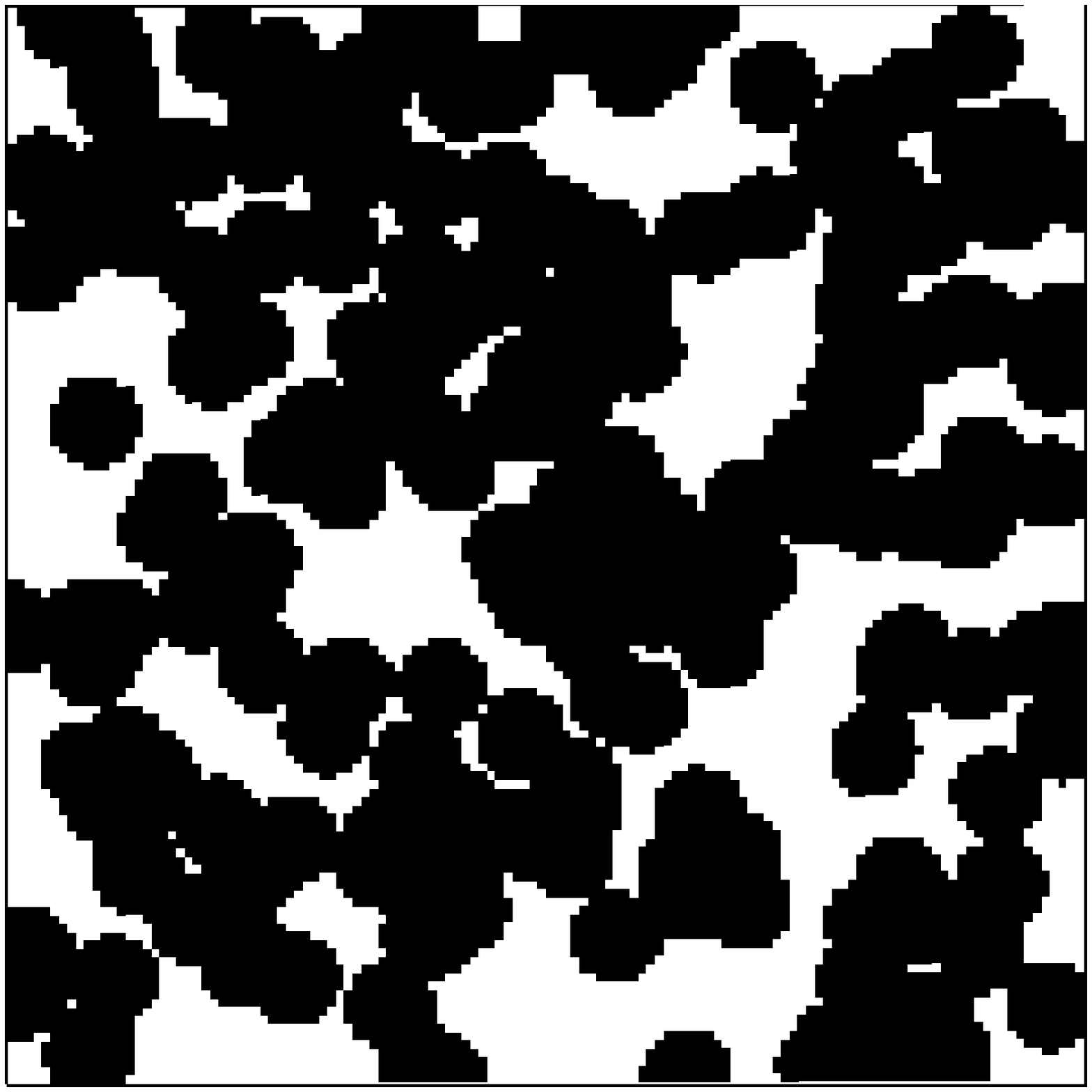}} \hfill
     \subfigure[Disconnected inclusion phase (low strength)]{
          \label{fig:ex11b}
             \includegraphics[width=.48\textwidth,height=5.5cm]{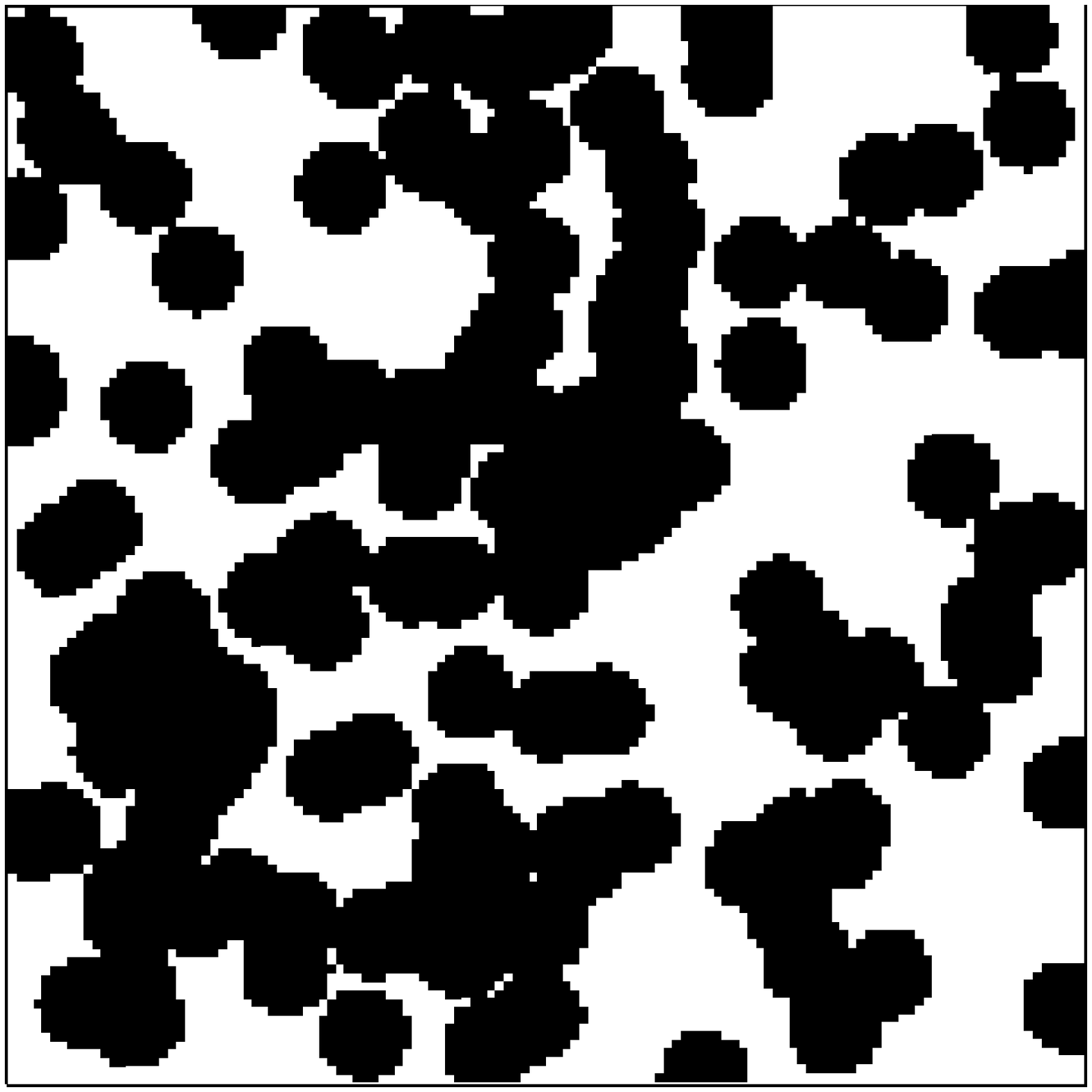}}\\
     \caption{Example 2}
     \label{fig:ex11}
\end{figure}

We consider a problem in nonlinear solid mechanics that illustrates the capabilities of the proposed methodology and the the significant improvements in computational efficiency even when very crude approximate solvers are selected. A random two-phase medium consisting of two elastic-perfectly-plastic materials occupies the unit square in 2D and is subjected to plane stress loading conditions. It was assumed that the {\em matrix} phase (white in Figure \ref{fig:ex11}) had a yield stress $\sigma_{yield}^{matrix}=0.1$ and the {\em inclusion} phase (black in Figure \ref{fig:ex11}) a yield stress $\sigma_{yield}^{inclusion}=1.0$ The same elastic properties were assumed for both phases (elastic modulus $E=1.0$ and Poisson's ratio $\nu=0.3$) and the von-Mises yield criterion was used. The stochasticity in the problem is introduced by the distribution of the black discs of diameter $d=0.0859375$ whose centers are assumed to follow a Poisson point process on the unit square. The intensity of the point process is selected so that the volume fraction of the inclusion  phase is $65\%$. This was intentionally chosen to be close to the percolation threshold of $68\%$ (\cite{yeo98rec}) in order to have realizations where the inclusion phase was connected and disconnected. In the former case, the load-bearing capacity (in the horizontal direction) of the specimen is high as the strong, inclusion phase forms a network that carries the load, whereas in the disconnected case, the load-bearing capacity is lower and determined by the yield stress of the matrix phase.
Hence the stochastic geometry completely determines the mechanical behavior of the system. The vector of uncertain parameters $\bxi$ ( \refeq{eq:def1}) consists of the number of inclusion disks  and the coordinates of their centers. As the former is a random variable (following a Poisson distribution)  the corresponding $\pi_{\xi}(\bxi)$ has support in spaces of varying dimension and non-zero probability for arbitrarily large $d$ where  $\bxi \in \RR^d$. It should be noted however that on average there are $181$ disks and  $dim(\bxi)=1+2~181=363$. Usual dimension reduction techniques based on second order properties  would be extremely misleading in this case as higher order statistics of the random medium (relating to connectivity) dominate mechanical response. 

The {\em exact} model  corresponds to a Finite Element solver with  $128 \times 128$ elements i.e. $11$ elements per inclusion diameter and $\approx 33,000$ dof, of the governing equations of equilibrium:
\be
\label{eq:prob1}
\nabla \cdot \bs{\hat{\sigma}} = 0
\ee
where $\bsf{\hat{\sigma}}$ is the Cauchy stress tensor and :
\be
\label{eq:prob2}
\dot{\hat{\sigma}}=\bs{c}: (\dot{\bs{\epsilon}}-\dot{\bs{\epsilon}}^p)
\ee
the constitutive rate equations with $\bs{\epsilon}$ and $\bs{\epsilon}^p$ being the total and plastic strain tensors respectively.

 The yield stress was assumed constant within each element and equal to the yield stress of the the phase occupying the majority of its area. The response of interest $y$ was the ultimate strength of the specimen in the horizontal direction and the average computational time was $\approx 700sec$ on a single CPU.

\begin{figure}[htp]
     \centering
     \subfigure[Exact model - $128 \times 128$ mesh ($700$ sec)]{
          \label{fig:ex12a}
           \includegraphics[width=.48\textwidth,height=5.5cm]{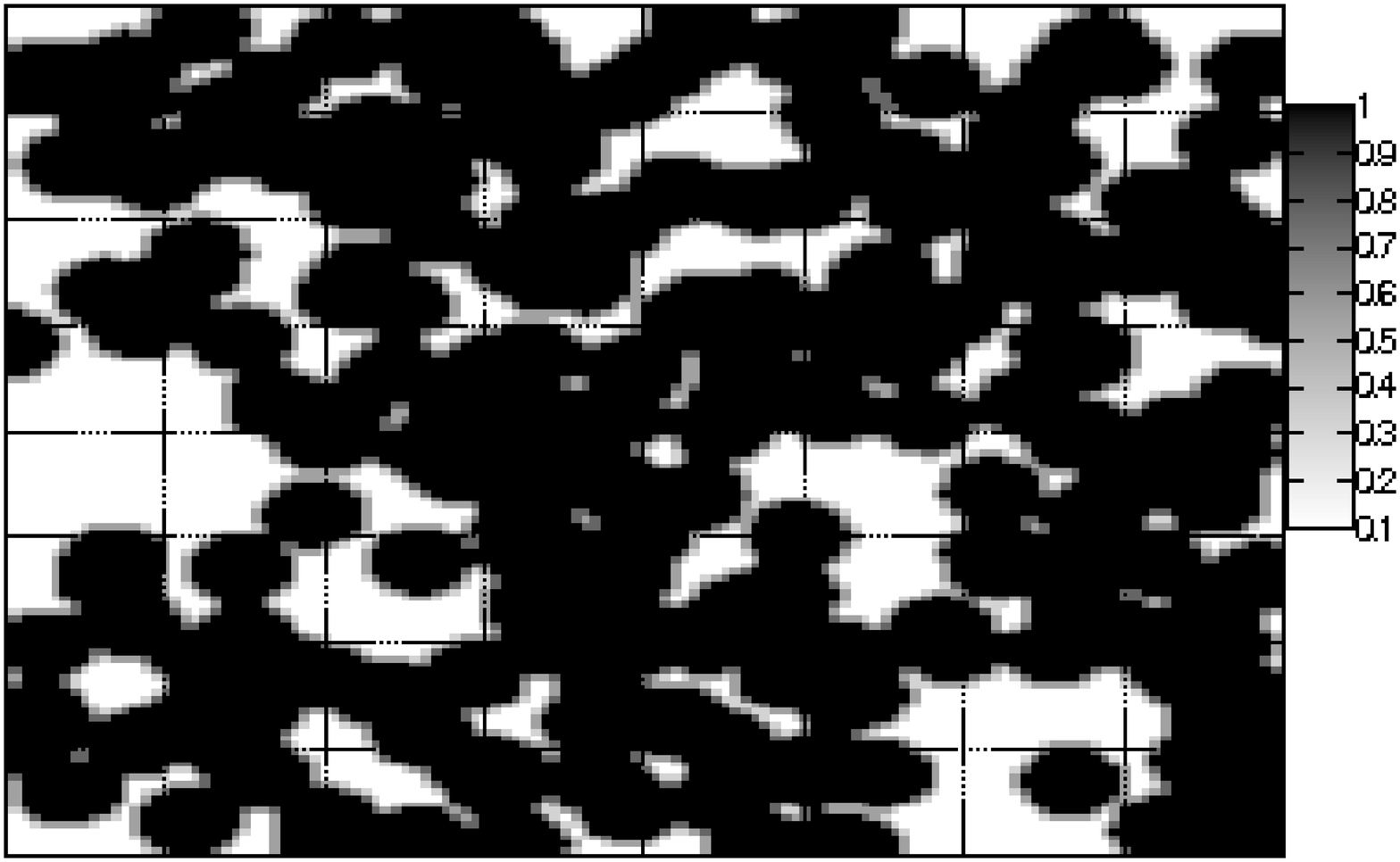}} \hfill
     \subfigure[Approximate model - $8 \times 8$ mesh ($0.15$ sec)]{
          \label{fig:ex12b}
             \includegraphics[width=.48\textwidth,height=5.5cm]{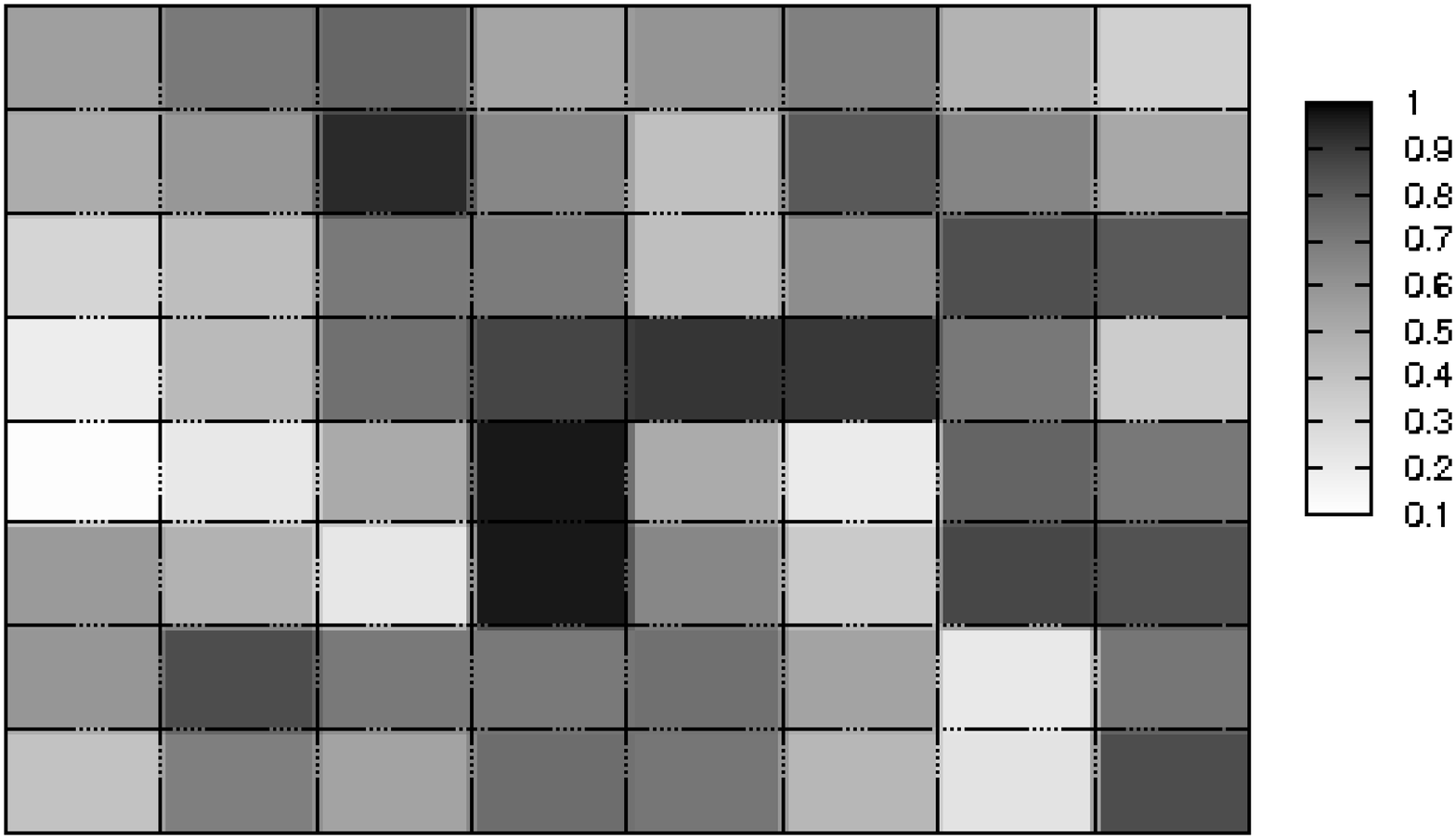}}\\
     \caption{Exact vs. Approximate solvers}
     \label{fig:ex12}
\end{figure}

\begin{figure}[htp]
\vspace{.5cm}
\psfrag{Y}{lalal}
\includegraphics[width=.8\textwidth,height=5.5cm]{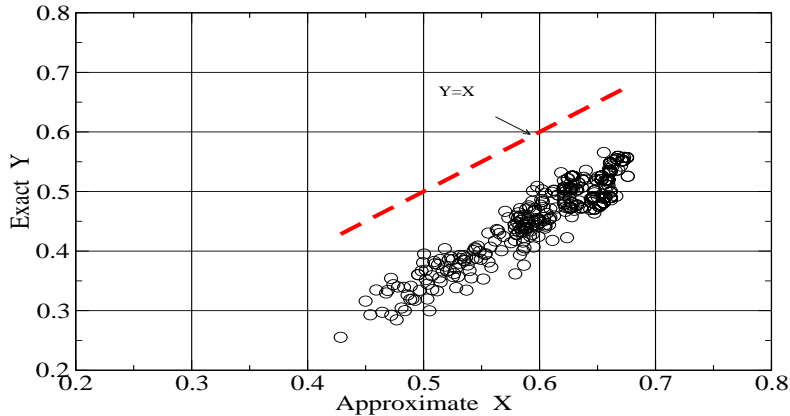}
\caption{Exact $y$ vs. Approximate $\bsf{x}$ response}
\label{fig:ex13}
\end{figure}

A single {\em approximate} model was used (i.e. $M=1$) which corresponds to Finite Element solver on a uniform $ 8 \times 8$  mesh and $128$ dof.  Constant yield stress was assigned to each of the $64$ elements based on the log-average of the yield stress within each element (Figure \ref{fig:ex12b}). Hence if $\mathcal{D}_e$ is the subdoman occupied by element $e$, its yield stress $\sigma_{yield,e}=\exp \{ \frac{1}{|\mathcal{D}_e |} \int_{\mathcal{D}_e} \log \left( \sigma_{yield}(\bs{s})\right)~d\bs{s} \}$.  The computational time for calculating the ultimate strength $\bsf{x}=x_1$ was $0.15$ sec, i.e. $\approx 4,700$ times faster than the {\em exact} model. It is obvious that such a solver introduces a significant error as it does not sufficiently resolve the governing PDEs and smears out the connectivity details that determine the ultimate strength of the specimen. Furthermore, the log-average rule used to determine the yield stress of the elements does not represent a consistent upscaling scheme of the material model. This discrepancy can be seen in Figure \ref{fig:ex13} which depicts $100$ pairs of approximate $\bsf{x}$  vs. exact response $y$. It is also observed that the mapping from $x_1$ to $y$ is one-to-many as the approximate model blurs some of the important microstructural details.

In order to compare the performance of the method we considered the  event $y > y_0=0.521$ which corresponds to a probability $10^{-3}$. This was found using an advanced simulation procedure  based on Sequential Monte Carlo as discussed in detail in (\cite{nea01ann,gel98sim,kou08des}).  The computational effort amounted to  $1,500$ calls to the {\em exact} solver and the coefficient of variation (c.o.v) of the estimate was $0.23$ (based on an asymptotic, and rather optimistic, bound in \cite{au01est}). This roughly implies that with probability $0.95$, the actual probability of the event of interest is in the interval $[0.55 \times 10^{-3},~ 1.5 \times 10^{-3}]$. It should be noted that to achieve the same c.o.v with standard Monte Carlo, $18,500$ calls to the {\em exact} solver would have been needed.

\begin{figure}
\psfrag{x0}{$x_0$}
\psfrag{px}{$Pr[x \ge x_0]$}
\includegraphics[width=.8\textwidth,height=5.5cm]{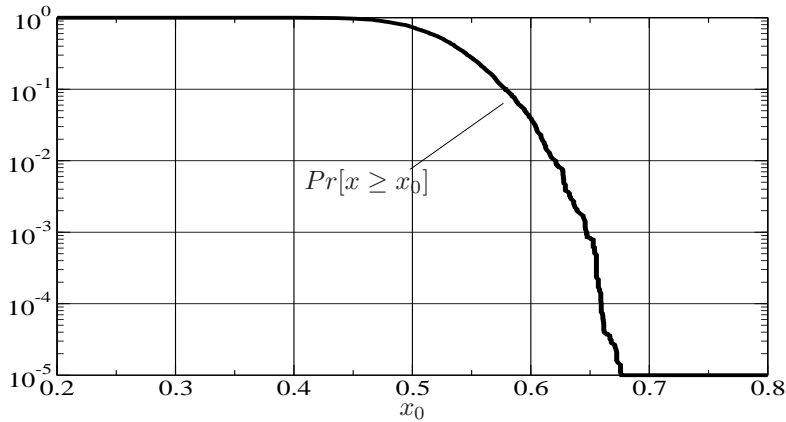}
\caption{Cumulative distribution function for the {\em approximate} solver output}
\label{fig:ex14}
\end{figure}

The density $\pi_{x}$ of $\bsf{x}=x_1$ was estimated using the same advanced Monte Carlo scheme that required $5,000$ calls to the {\em approximate} solver. The corresponding cdf is depicted in Figure \ref{fig:ex14} for probabilities as low as $10^{-5}$. Note that due to the reduced computational effort associated with calls to the approximate solver, the effective computational time for this task amounted to (approximately) $1$ call to the {\em exact} solver.

As with the previous example  Figure \ref{fig:ex15} depicts posterior statistics of this model for various training sample sizes. Figure \ref{fig:ex16} depicts the posterior mean and posterior quantiles of $q_{\mathcal{A}}(\bsf{x})$ based on Equations (\ref{eq:def1e}) and (\ref{eq:def5}) for various sample sizes. Table \ref{tab:ex11} summarizes the estimates based on the posterior mean $\hat{q}_{\mathcal{A}}(\bsf{x})$  and confidence bounds established with  $q_{\mathcal{A},0.01}(\bsf{x})$ and $ q_{\mathcal{A},0.99}(\bsf{x})$.
 It is noted that good estimates of the actual output statistic can be obtained at a fraction of the computational cost. Furthermore good confidence bounds are readily obtained for all sample sizes. The same good quality in the results is also observed in the estimates of the whole cdf of the exact output which is depicted in Figure \ref{fig:ex17} for $100$ training samples.




\begin{figure}[htp]
     \centering
     \subfigure[$10$ training samples]{
          \label{fig:ex15a}
           \includegraphics[width=.48\textwidth,height=3.5cm]{FIGURES/samples_exact-approximate_010.eps}} \hfill
     \subfigure[$10$ training samples]{
          \label{fig:ex15b}
\psfrag{sigma}{$\sigma$}
             \includegraphics[width=.48\textwidth,height=3.5cm]{FIGURES/posterior_variance_010.eps}}\\
\vspace{.2cm}
     \subfigure[$20$ training samples]{
          \label{fig:ex15c}
           \includegraphics[width=.48\textwidth,height=3.5cm]{FIGURES/samples_exact-approximate_020.eps}} \hfill
     \subfigure[$20$ training samples]{
          \label{fig:ex15d}
\psfrag{sigma}{$\sigma$}
             \includegraphics[width=.48\textwidth,height=3.5cm]{FIGURES/posterior_variance_020.eps}}\\
\vspace{.2cm}
     \subfigure[$50$ training samples]{
          \label{fig:ex15e}
           \includegraphics[width=.48\textwidth,height=3.5cm]{FIGURES/samples_exact-approximate_050.eps}} \hfill
     \subfigure[$50$ training samples]{
          \label{fig:ex15f}
\psfrag{sigma}{$\sigma$}
             \includegraphics[width=.48\textwidth,height=3.5cm]{FIGURES/posterior_variance_050.eps}}\\
\vspace{.2cm}
     \subfigure[$100$ training samples ]{
          \label{fig:ex15g}
           \includegraphics[width=.48\textwidth,height=3.5cm]{FIGURES/samples_exact-approximate_100.eps}} \hfill
     \subfigure[$100$ training samples]{
          \label{fig:ex15h}
\psfrag{sigma}{$\sigma$}
             \includegraphics[width=.48\textwidth,height=3.5cm]{FIGURES/posterior_variance_100.eps}}\\
     \caption{Posterior mean of $f(\bsf{x}; \bt)$ and posterior density $\sigma$ for various training sample sizes}
     \label{fig:ex15}
\end{figure}

\begin{figure}[htp]
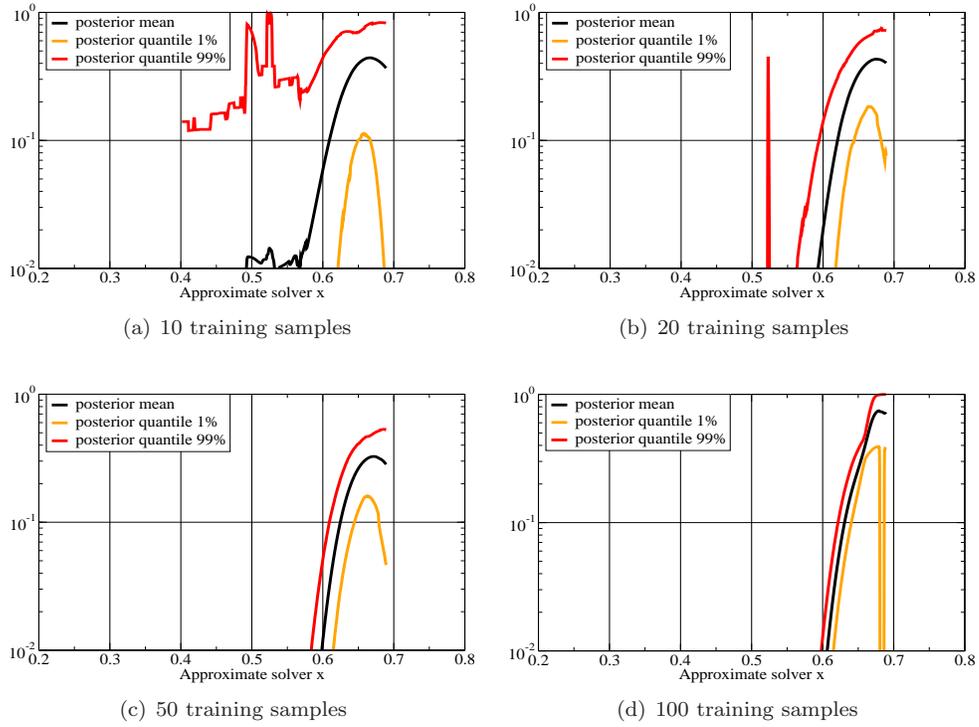

     \centering
     \subfigure[$10$ training samples]{
          \label{fig:ex16a}
           \includegraphics[width=.48\textwidth,height=4.cm]{FIGURES/conditional_0010.eps}} \hfill
     \subfigure[$20$ training samples ]{
          \label{fig:ex16b}
\psfrag{sigma}{$\sigma$}
             \includegraphics[width=.48\textwidth,height=4.cm]{FIGURES/conditional_0020.eps}}\\
\vspace{.2cm}
     \subfigure[$50$ training samples ]{
          \label{fig:ex16c}
           \includegraphics[width=.48\textwidth,height=4.cm]{FIGURES/conditional_0050.eps}} \hfill
     \subfigure[$100$ training samples]{
          \label{fig:ex16d}
\psfrag{sigma}{$\sigma$}
             \includegraphics[width=.48\textwidth,height=4.cm]{FIGURES/conditional_0200.eps}}\\

     \caption{Posterior mean and quantiles for $Pr[ y \ge 0.521 \mid x]$ based on various sample sizes}
     \label{fig:ex16}
\end{figure}

\begin{table}[htp]
\caption{Estimates of $Pr[ y > y_0=0.521]$ and computational effort (the latter is measured in number equivalent number of calls to the {\em exact} solver}
\begin{center}
\begin{tabular}{c|c|c|c|c|}
\hline
Number & Posterior & Posterior & Posterior & Computational \\
of samples & mean  & quantile $1\%$ & quantile $99\%$ & Effort\\
\hline \hline
 10 &  $1.47\times 10^{-2}$ &   $2.33 \times 10^{-4}$ & $3.80 \times 10^{-1}$ &  $11$  \\
\hline
 20 & $6.24 \times 10^{-3}$ &  $3.56 \times 10^{-3}$  & $1.90 \times 10^{-2}$   & $21$ \\
\hline
 30 & $ 2.64 \times 10^{-3}$  &  $3.50 \times 10^{-4}$ &   $8.55 \times 10^{-3}$ & $31$  \\
\hline
50 &  $ 2.64   \times 10^{-3}$  &  $4.25 \times 10^{-4}$   & $ 5.23 \times 10^{-3}$ & $51$ \\
\hline
 100 &  $1.06 \times 10^{-3}$ &   $4.75 \times 10^{-4}$ & $2.14 \times 10^{-3}$  & $101$ \\
\hline
\end{tabular}
\end{center}
\label{tab:ex11}
\end{table}

\begin{figure}
\vspace{.5cm}
\psfrag{y0}{$y_0$}
\psfrag{py}{$Pr[ y \ge y_0]$}
\includegraphics[height=4.5cm,width=0.8\textwidth]{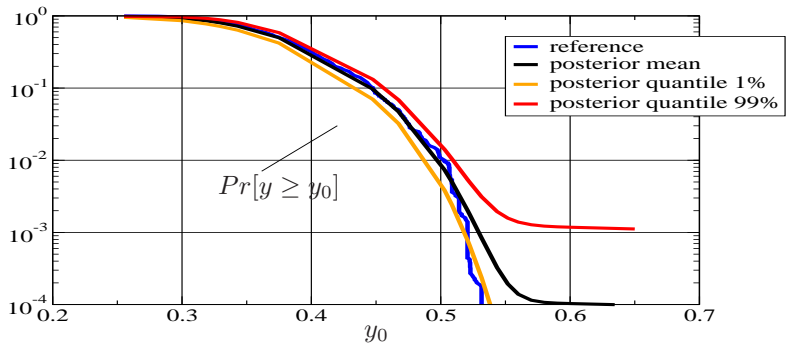}
\caption{Posterior for $Pr[ y \ge y_0] ~\forall y_0$ based on 100 training samples}
\label{fig:ex17}
\end{figure}


We examined the same problem using $M=2$ approximate solvers.  In addition to the approximate solver discussed earlier we considered a second FE model with the same mesh of $64$ elements (Figure \ref{fig:ex12b}). The yield stress $\sigma_{yield,e}$ for each element $e$ was now determined by averaging i.e. $\sigma_{yield,e}= \frac{1}{|\mathcal{D}_e |} \int_{\mathcal{D}_e}  \sigma_{yield}(\bs{s})~d\bs{s} $. This is again a non-consistent rule with the exact constitutive model and would yield approximate solutions. Furthermore due to the concavity of the log-function and Jensen's inequality, the assigned yield stresses $\sigma_{yield,e}$ were smaller in the first model and as a result the predicted approximate outputs $x_1 < x_2$. 

\begin{figure}
\centering
\includegraphics[width=.8\textwidth,height=6cm]{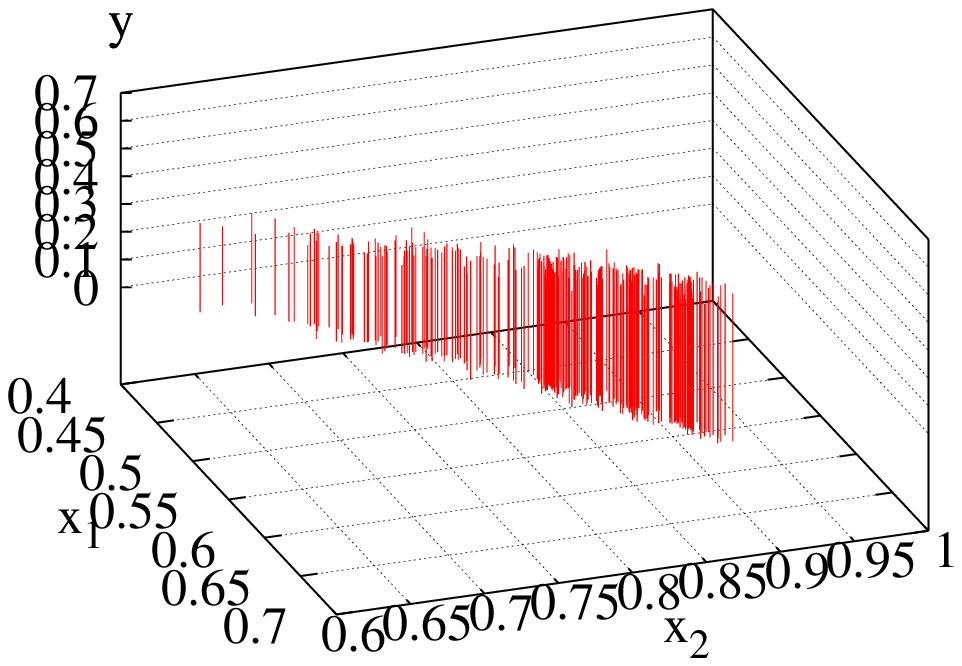}
\caption{Exact $y$ vs. Approximate $\bsf{x}=(x_1,x_2)$ response}
\label{fig:ex18}
\end{figure}

Figure \ref{fig:ex18} depicts $50$ triplets $(x_1,x_2,y)$ comparing the exact output with the approximate solutions provided by the two reduced models. It is expected that the addition of the second model will yield more information about $y$ that can be readily taken into account by the Bayesian framework presented. The joint pdf $\pi_{\bs{x}}(x_1,x_2)$ was estimated using $5,000$ calls to each solver (i.e. total $10,000$) which due to their reduced computational cost amounted to the equivalent of $\approx 2$ runs of the exact solver.

\begin{figure}[htp]
     \centering
     \subfigure[ ]{
          \label{fig:ex19a}
           \includegraphics[width=.48\textwidth,height=4.cm]{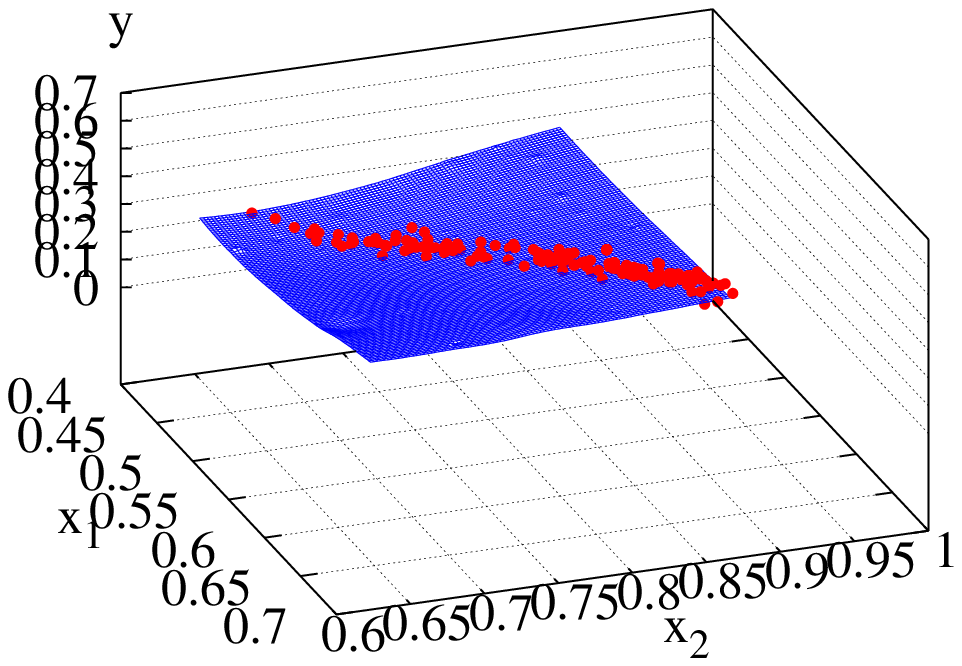}} \hfill
     \subfigure[ ]{
          \label{fig:ex19b}
\psfrag{sigma}{$\sigma$}
             \includegraphics[width=.48\textwidth,height=4.cm]{FIGURES/boolean2_posterior_sigma_0050.eps}}\\
     \caption{Posterior mean of $f(\bsf{x}; \bt)$ and posterior density $\sigma$ for $50$ training samples}
     \label{fig:ex19}
\end{figure}

\begin{figure}
\centering
\psfrag{y0}{$y_0$}
\psfrag{py}{$Pr[ y \ge y_0]$}
\includegraphics[height=4.5cm,width=0.8\textwidth]{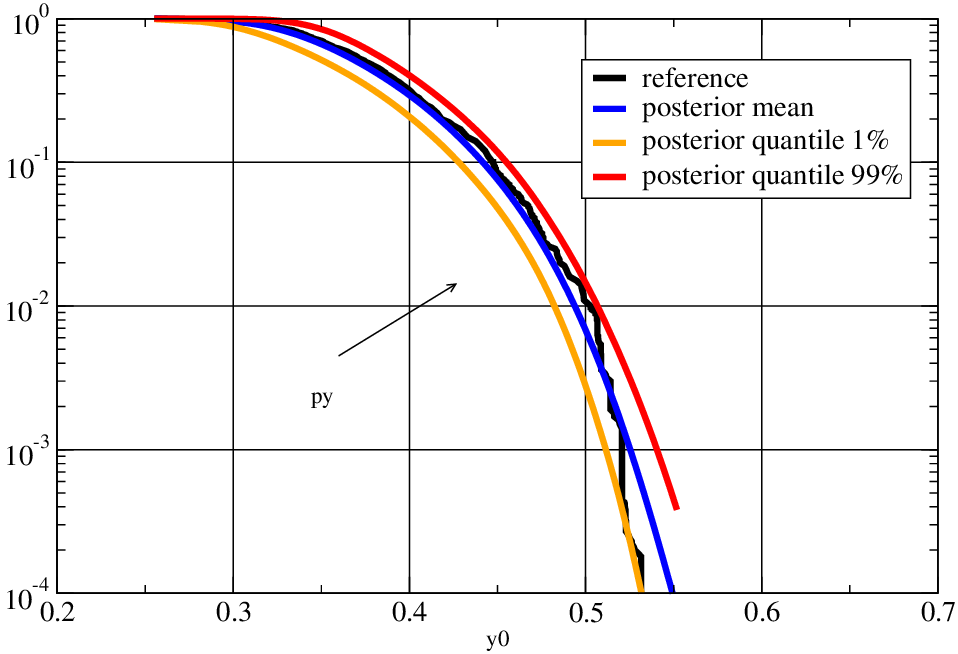}
\caption{Posterior mean and quantiles for $Pr[ y \ge y_0 ], ~ \forall y_0$ based on $50$ training samples}
\label{fig:ex110}
\end{figure}

Figure \ref{fig:ex19} depicts posterior statistics of the regression model for  50 training samples. Finally Figure \ref{fig:ex110} illustrates the complete cdf of the exact output $y$ based on the same number of training samples. The computational effort for obtaining this result is equivalent to $52$ calls to the exact solver (i.e. $2$ for estimating $\pi_{x}$ and $50$ for obtaining the training data. The addition of the second predictor $x_2$ offers a significant improvement w.r.t. Figure \ref{fig:ex17} not only in terms of accuracy but also in terms computational efficiency since the latter result required effectively $101$ calls to the exact solver.

\section{Conclusions}
The majority of systems of physical and engineering interest are characterized by a large number of uncertainties. These are non-Gaussianly distributed and quite frequently their higher-order properties play a decisive role in the statistics of the response/output. While Monte Carlo techniques provide the only general  method for uncertainty quantification in such systems and  despite the significant progress of recent years, they might still require an infeasible number of calls to the forward solver.
The present paper introduced a Bayesian  framework where outputs from approximate, inexpensive  solvers can be rigorously utilized in order to accelerate the solution process. We made use of a flexible, non-parametric Bayesian model and a general SMC-based inference engine that is able to establish a quantitative link between approximate and exact solver. This can in turn be used to produce estimates for the output statistics of interest and rigorous confidence bounds. 
While this capability was not utilized in the examples presented, these credible intervals can assist  in  minimizing  the number of calls to the exact solver by performing  runs in selected regions that will be most informative. Furthermore it offers the capability of utilizing multiple approximate solvers and it opens the door for designing or optimizing systems in the presence of uncertainties based on establishing functions of the output statistics with respect to the design variables.






\bibliographystyle{siam}
\bibliography{paper}

\end{document}